\documentclass[%
reprint,
 amsmath,amssymb,
 aps,
]{revtex4-2}

\usepackage{svg}
\usepackage{tikz}
\usetikzlibrary{quantikz}
\usepackage{braket}
\usepackage{lipsum}
\usepackage{graphicx}% Include figure files
\usepackage{dcolumn}% Align table columns on decimal point
\usepackage{bm}% bold math
\usepackage{physics}
\usepackage{hyperref}
\usepackage{url}
\usepackage{xcolor}
\usepackage{bbm}

\usepackage{amsthm}

\newcommand{\rvline}{\hspace*{-\arraycolsep}\vline\hspace*{-\arraycolsep}}

\theoremstyle{plain}
\newtheorem{theorem}{Theorem}

\newtheorem{proposition}[theorem]{Proposition}
\newtheorem{conjecture}[theorem]{Conjecture}
\newtheorem{corollary}[theorem]{Corollary}
\newtheorem{definition}{Definition}

\newcommand{\Mod}[1]{\ (\mathrm{mod}\ #1)}

\def\checkmark{\tikz\fill[scale=0.4](0,.35) -- (.25,0) -- (1,.7) -- (.25,.15) -- cycle;} 

\begin{document}

\preprint{APS/123-QED}

\title{Generalized Concentratable Entanglement via Parallelized Permutation Tests}% Force line 

\author{Xiaoyu Liu$^{1}$}%
\author{Johannes Knörzer$^{2}$}%
\author{Zherui Jerry Wang$^{1,3}$}%
\author{Jordi Tura$^{1}$}%
\affiliation{$^1$Instituut-Lorentz, Universiteit Leiden, P.O. Box 9506, 2300 RA Leiden, The Netherlands}%
\affiliation{$^2$Institute for Theoretical Studies, ETH Zurich, 8092 Zurich, Switzerland}%
\affiliation{$^3$Okinawa Institute of Science and Technology Graduate University, Onna-son, Okinawa 904-0495, Japan}

\date{\today}% It is always \today, today,

\begin{abstract}
Multipartite entanglement is an essential resource for quantum information theory and technologies, but its quantification has been a persistent challenge.
Recently, \textit{Concentratable Entanglement} (CE) has been introduced as a promising candidate for a multipartite entanglement measure, which can be efficiently estimated across two state copies.
In this work, we introduce \textit{Generalized Concentratable Entanglement} (GCE) measures, highlight a natural correspondence to quantum Tsallis entropies, and conjecture a new entropic inequality that may be of independent interest.
We show how to efficiently measure the GCE in a quantum computer, using parallelized permutation tests across a prime number of state copies.
We exemplify the practicality of such computation for probabilistic entanglement concentration into $\ket{W}$ states with three state copies.
Moreover, we show that an increased number of state copies provides an improved error bound on this family of multipartite entanglement measures in the presence of imperfections.
Finally, we prove that GCE is still a well-defined entanglement monotone as its value, on average, does not increase under local operations and classical communication (LOCC).
\end{abstract}

\maketitle

%\tableofcontents

% \section{\label{intro}Introduction}

\textit{Introduction.}\textemdash
Quantum entanglement is one of the most intriguing and fundamental phenomena in quantum mechanics~\cite{einstein1935can}.
As a valuable resource in quantum information processing, entanglement is crucial for quantum networks~\cite{munro2015quantum, wehner2018quantum,zhong2021deterministic}, distributed quantum computing~\cite{buhrman2003distributed,cirac1999distributed,schiffer2023quantum} and quantum sensing~\cite{degen2017quantum,zhang2015entanglementenhanced,zhuang2018distributed}.
To certify the functionality of these applications, it is important to verify and quantify the degree of entanglement in quantum systems~\cite{elben2020crossplatform, knorzer2023crossplatform, zhang2022certification}.
However, large-scale quantum networks and computers include numerous quantum subsystems, making this characterization challenging.
Several multipartite entanglement measurement techniques have been proposed~\cite{coffman2000distributed,barnum2001monotones,wong2001potential,meyer2002global,walter2013entanglement,brennen2003observable,carvalho2004decoherence}, but they are often impractical to estimate and limited by system size.

Recently, a family of multipartite entanglement measures called \textit{Concentratable Entanglement} (CE) has been proposed~\cite{beckey2021computable,beckey2023multipartite,coffman2024local}. 
Mathematically, it is the arithmetic mean of the linear entropy (the first order approximation of von Neumann entropy), $1-\Tr(\rho^2_{\alpha})$~\cite{vonneumann2018mathematical,santos2000linear}, of all possible subsystems $\alpha$. In some special cases, CE can recover several well-known entanglement measures~\cite{meyer2002global,wong2001potential,walter2013entanglement,brennen2003observable,carvalho2004decoherence}. 
Moreover, it can be estimated efficiently via parallelized SWAP tests between two copies of a given quantum state $\ket{\psi}$~\cite{barenco1997stabilization,buhrman2001quantum,foulds2021controlled}.
As collective measurements on multiple copies of a quantum system are known to be advantageous for single-system property testing~\cite{cotler2019quantum,koczor2021exponential,huggins2021virtual}, it is compelling to consider the generalization of CE to more than two copies of $\ket{\psi}$, both mathematically and practically. 

In this work, we introduce \textit{Generalized Concentratable Entanglement} (GCE) and reveal its close relation to quantum Tsallis entropies $T_K(\rho)$~\cite{tsallis1988possible,caruso2008nonadditive}, for any real $K>1$. 
For any prime number $K$, we prove that GCE can be efficiently measured in a quantum computer using $K$ copies of a state $\ket{\psi}$ and a parallelized permutation test~\cite{kada2008efficiency,buhrman2024permutation}. 
We demonstrate that, up to local unitaries, $\ket{W}$ states can be probabilistically extracted from the permutation test when $K=3$, thereby concentrating the entanglement into $\ket{W}$ states. 
Furthermore, we analyze the errors in the estimated GCE for a constant number of noisy input states and show that they decrease with the number of state copies as $O(\frac{1}{K})$. 
We also prove that GCE is still a well-defined entanglement monotone as its value, on average, does not increase under local operations and classical communication (LOCCs). 
In the end, we present several mathematical properties of GCE and, supported by strong numerical evidence, present two conjectures, which may provide deeper mathematical insights into the features of both GCE and Tsallis entropies.

\bigskip

\textit{Generalized Concentratable Entanglement.}\textemdash
We consider the following definition of GCE. 

\begin{definition}[\textit{Generalized Concentratable Entanglement}]
Consider an input $n$-qubit pure state $\ket{\psi}$ with labels $S=\{1,2,\cdots,n\}$ for each qubit respectively, and one measures the entanglement of every non-empty subsystem $s$ in the power set of $S$, i.e., $s\in\mathcal{P}(S)\backslash\{\varnothing\}$. 
The Generalized Concentratable Entanglement (GCE) is defined as:

\begin{equation}
    \mathcal{C}^{(K)}_{\ket{\psi}}(s):=\frac{1}{K-1}\left(1-\frac{1}{2^{|s|}}\sum_{\alpha\in\mathcal{P}(s)}\Tr(\rho_{\alpha}^K)\right),
    \label{defeq}
\end{equation}
for any $K>1$. 
Here, $|s|$ is the cardinality of $s$. 
$\rho_{\alpha}$ denotes the corresponding reduced density matrix of subsystem $\alpha\in\mathcal{P}(s)$ where $\rho=\ketbra{\psi}$. 
We take $\Tr(\rho^K_{\varnothing})=1$ in the sum. 

\end{definition}

Equivalently, the definition of $\mathcal{C}^{(K)}_{\ket{\psi}}(s)$ can be viewed as the arithmetic mean of the Tsallis entropies:
\begin{equation}
    T_K(\rho_{\alpha})=\frac{1}{K-1}\left( 1-\Tr(\rho_{\alpha}^K)  \right),
\end{equation}
for all subsystems $\alpha$ of the measured system $s$. 
Notably, when $K=2$, Eq.~\eqref{defeq} recovers the original CE defined in~\cite{beckey2021computable}. 
Moreover, when $K\rightarrow1$, Eq.~\eqref{defeq} becomes the arithmetic mean of the von Neumann entropy~\cite{vonneumann2018mathematical}.

\bigskip

\textit{Efficient Estimation of GCE.}\textemdash
Eq.~\eqref{defeq} includes the term $\Tr(\rho^K_{\alpha})$. 
When $K$ is a positive integer, it can be estimated via a controlled-derangement operator $D$ decomposed in controlled-SWAP operations~\cite{koczor2021exponential,quek2024multivariate,huggins2021virtual}. 
One possible choice for $D$ is schematically given by:

\begin{equation}
\small
    \begin{quantikz}[row sep = {0.5cm,between origins}]
    \lstick{1} & \gate[5]{D} &\qw \rstick{2}\\
    \lstick{2} &  &\qw \rstick{3}\\
    \lstick{3} &  &\qw \rstick{4}\\
    \lstick{\vdots} & &\qw  \rstick{\vdots}\\
    \lstick{$K$} & &\qw \rstick{1}\\
    \end{quantikz}=
    \begin{quantikz}[row sep = {0.5cm,between origins}]
    \lstick{1} & \swap{1} &\qw &\qw &\qw &\qw \rstick{2} \\
    \lstick{2} & \targX{} & \swap{1} &\qw &\qw &\qw\rstick{3} \\
    \lstick{3} &\qw & \targX{} & \swap{1} &\qw &\qw\rstick{4} \\
    \lstick{\vdots} &\qw &\qw & \targX{} & \swap{1} &\qw\rstick{\vdots}  \\
    \lstick{$K$} &\qw &\qw &\qw & \targX{} &\qw\rstick{1}  \\
    \end{quantikz},
    \label{D1}
\end{equation}
which shuffles the system positions from $\{1,2,\cdots,K-1,K\}$ into $\{2,3,\cdots,K,1\}$. 
We remark, in order to estimate $\Tr(\rho^K_{\alpha})$ in this way, one needs to prepare $K$ copies of $\rho_{\alpha}$ for each $\alpha$.

It was shown that for $K=2$ the GCE can be estimated efficiently via a series of parallelized SWAP tests. 
While an extension to multiple copies via a series of parallelized derangement operators might seem plausible, such an approach is not viable, since in general $D$ is not Hermitian.

\begin{figure}
    \centering
    \includegraphics[width=1.0\linewidth]{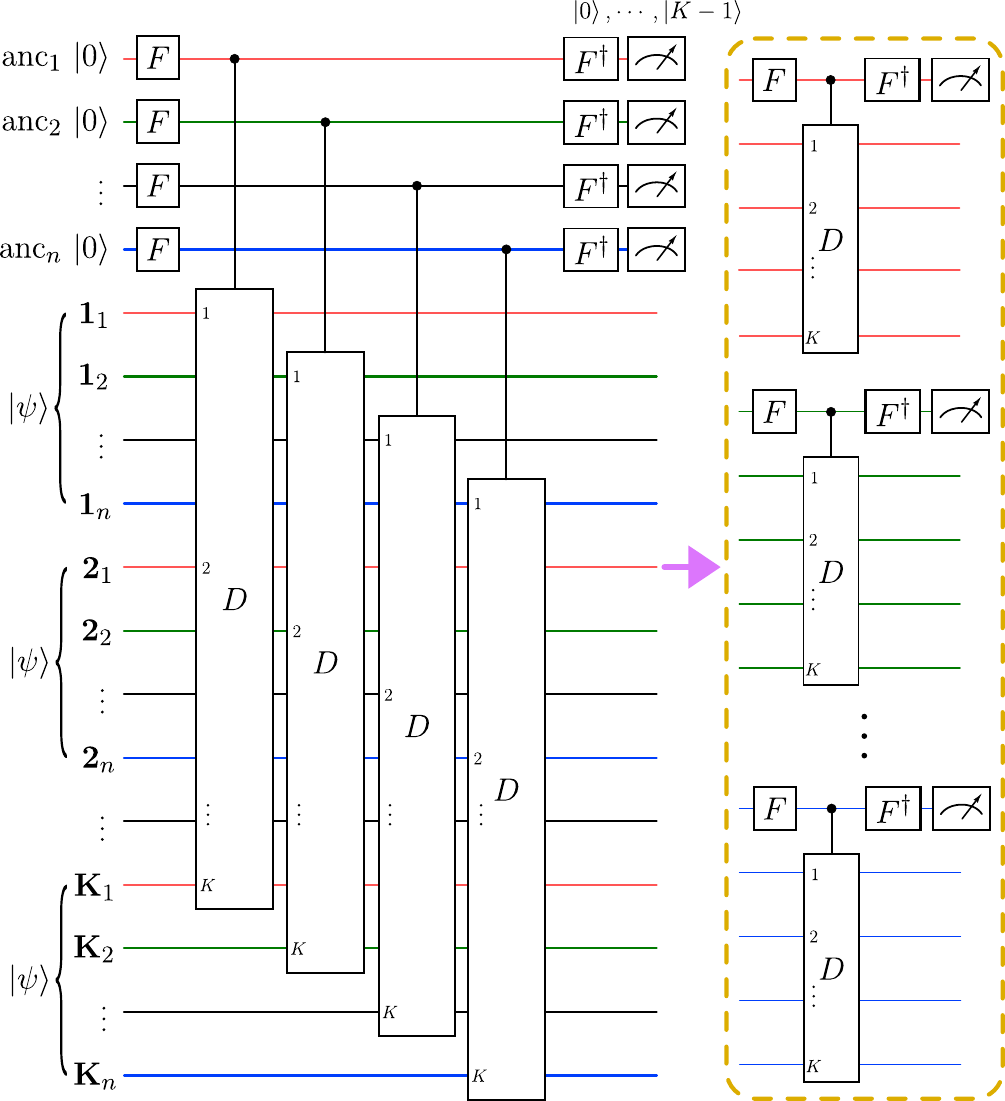}
    \caption{Circuit structure for computing GCE with prime $K$. 
    Multiple copies of state $\ket{\psi}$ are prepared and $n$ permutation tests are performed on the state copies in parallel with the help of $n$ $K$-level qudit ancillas. 
    The probability distribution of the measurement result on the ancillas can be efficiently sampled, simply by running the circuit, and hence one can estimate the GCE via Eq.~\eqref{eqcecompute}. 
    Notably, this only holds for any prime $K$. The analytical proof for this can be found in Appendix~\ref{proofcircuit}.}
    \label{circfig}
\end{figure}

Here we propose a quantum circuit building on the parallelized permutation test with $K$-level ancillas~\cite{kada2008efficiency,buhrman2024permutation} to efficiently estimate the GCE for any prime $K$. 
The corresponding circuit is shown in Fig.~\ref{circfig}. 
To begin with, $K$ copies of $\ket{\psi}$ are prepared and $n$ ancillary $K$-level qudits are initialized in $\ket{0}$. 
Then, all ancillas are acted upon by a quantum Fourier transform $F: \ \ket{z}\rightarrow\frac{1}{\sqrt{K}}\sum_{k=0}^{K-1}\omega^{zk}\ket{k}$ where $\omega=e^{2\pi i/K}$. 
Subsequently a multi-level controlled-$D$ operator is applied in parallel to qubits in each copy that share the same label.
Its action can be schematically written as:
\begin{equation}
    \begin{quantikz}[row sep = {0.5cm, between origins}]
    \lstick{$\sum_{z=0}^{K-1}c_z\ket{z}$}& \ctrl{1} &\qw \\
    \lstick{$\ket{\phi}$}&\gate{D} &\qw \\
    \end{quantikz} \ = \ \sum_{z=0}^{K-1} c_z \ket{z} \otimes D^z \ket{\phi},
    \label{Dops}
\end{equation}
with $c_z\in\mathbb{C}$ and $\sum_{z=0}^{K-1}|c_z|^2=1$. 
Finally, the inverse Fourier transforms $F^{\dagger}$ are applied on each ancilla and one measures them eventually. 
By running the circuit with sufficiently many repetitions, one can obtain the probability distribution of the resulting digit strings on the ancillas, \textit{i.e.}, $p(\mathbf{z})=p(z_1 z_2 \cdots z_n)$, where $\mathbf{z}\in\{0,1,\cdots,K-1\}^{n}$. 
From $p(\mathbf{z})$, one is able to estimate the GCE in Eq.~\eqref{defeq} for prime $K$. The reason why this estimation procedure only works for prime $K$ is related to the existence of single-cyclic permutations for arbitrary powers of $D$, as is detailed in Appendix~\ref{proofcircuit}.

\begin{figure*}
    \centering
    \includegraphics[width=1.0\linewidth]{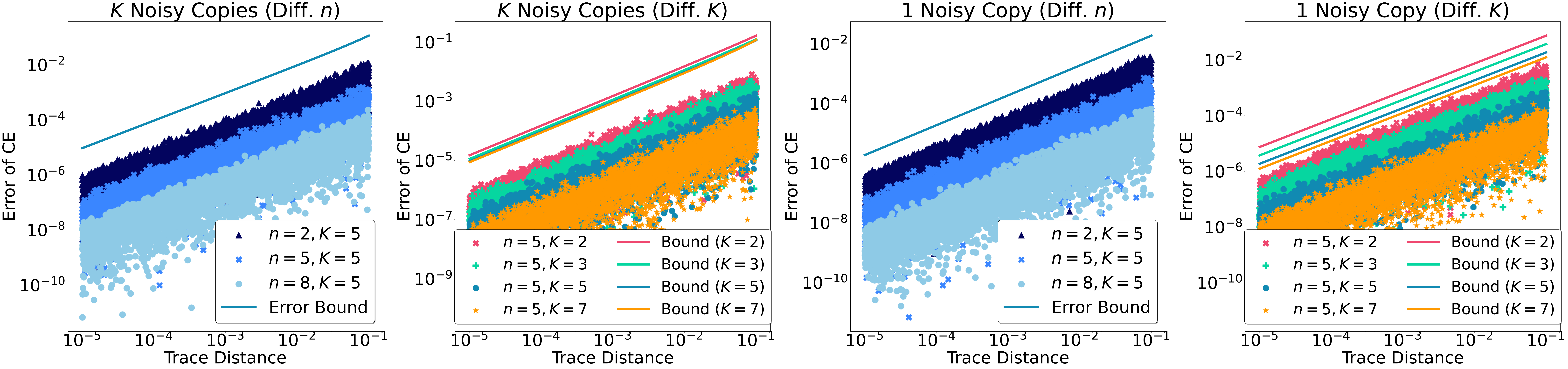}
    \caption{Numerical simulations for GCE errors. 
    We have considered 5000 samples of Haar-random input states $\ket{\psi}$ for each $\{K,n\}$ and $|s|=2$ for all cases. 
    The leftmost two plots show scenario 1, where there are $K$ noisy copies of $\ket{\psi}$, each with trace distance $\epsilon$ to the perfect state. 
    The rightmost two plots show scenario 2, where there is only 1 noisy copy. One can see that all the errors are upper-bounded by Eq.~\eqref{errbound}. 
    Also, on average, the GCE errors decrease for larger $K$ and $n$. 
    Our study suggests there may exist sharper upper-bounds.}
    \label{errorfig}
\end{figure*}

\begin{proposition}\label{prop1}
Sampling from the probability distribution $p(\mathbf{z})$, we can estimate $\Tr(\rho_{\alpha}^K)$ for any subsystem $\alpha$ of the pure state $\ket{\psi}$ via two equivalent ways:
\begin{equation}
\begin{split}
    \Tr(\rho^K_{\alpha}) =& \frac{1}{K-1}\left(K\sum_{\sum_{x\in\alpha}z_x\equiv 0 \Mod{K}}p(\mathbf{z})-1\right) \\
    =& 1- K\sum_{\sum_{x\in\alpha}z_x\not\equiv 0 \Mod{K}}p(\mathbf{z}).
\end{split}
\end{equation}
As $\Tr(\rho_{\alpha}^K)$ are included in Eq.~\eqref{defeq}, we can then naturally estimate GCE as:
\begin{equation}
\begin{split}
    &\mathcal{C}^{(K)}_{\ket{\psi}}(s) \\
    =&\frac{K}{2^{|s|}(K-1)^2}\sum_{\alpha\in\mathcal{P}(s)}\left(1-\sum_{\sum_{x\in\alpha}z_x\equiv 0 \Mod{K}}p(\mathbf{z})\right) \\
    =&\frac{K}{2^{|s|}(K-1)}\sum_{\alpha\in\mathcal{P}(s)}\sum_{\sum_{x\in\alpha}z_x\not\equiv 0 \Mod{K}}p(\mathbf{z}).
\end{split}
\label{eqcecompute}
\end{equation}
    
\end{proposition}

From Proposition \ref{prop1}, one can find that:
\begin{corollary}\label{corollary}
$p(\mathbf{z'})=0$ for any $h(\mathbf{z'})\not\equiv0 \Mod{K}$, where $h(\mathbf{z'})$ denotes the sum over the digits in the string $\mathbf{z'}=z'_1z'_2\cdots z'_n$, i.e., $h(\mathbf{z'})=\sum_{j=1}^{n}z'_j$.
\end{corollary}

Notably, when $K=2$, the circuit in Fig. \ref{circfig} recovers the parallelized SWAP test from Ref.~\cite{beckey2021computable}, as the Fourier transforms become Hadamard gates and controlled-$D$ operators become Fredkin gates.
In addition, the circuit in Fig.~\ref{circfig} can be encoded into a qubit system, possibly with redundant computing power, which may be more suitable for practical experiments.
Moreover, this approach does not exactly hold for composite $K$, which is also explained in Appendix~\ref{proofcircuit}. 

\bigskip

\textit{Entanglement Concentration for $K=3$.}\textemdash
When $K=2$, there is a probabilistic way to concentrate the entanglement (Bell pairs) from the input states at the controlled system of each parallelized SWAP test. 
In this work, we provide a similar proposition for $K=3$:

\begin{proposition}\label{prop2}
For $K=3$, there is a probabilistic entanglement concentration for each parallelized permutation test. 
Specifically, for each parallelized permutation test, when the top ancilla clicks at either $\ket{1}$ or $\ket{2}$, there exists a set of local unitaries that converts the controlled system into $\ket{W}$ states, i.e., 
\begin{equation}
\small
    \begin{quantikz}[row sep = {0.5cm, between origins}]
    \lstick{$\ket{0}$} & \gate{F} &\ctrl{1}  &\gate{F^{\dag}} &\meter{$\ket{1},\ket{2}$} \\
    \lstick{$\mathbf{1}_j$} &\qw &\gate[3]{D}  &\qw&\qw \rstick[3]{$\left(U_{\mathbf{1}_j}\otimes U_{\mathbf{2}_j} \otimes U_{\mathbf{3}_j}\right)^{\dagger} \ket{W}$}  \\
    \lstick{$\mathbf{2}_j$} &\qw  &\qw&\qw&\qw  \\
    \lstick{$\mathbf{3}_j$} &\qw  &\qw&\qw&\qw \\
    \end{quantikz}
\end{equation}
for $\forall j\in\{1,2,\cdots,n\}$. 
\end{proposition}

For $K>3$, there does not in general exist a set of local unitaries that convert the controlled system into either $\ket{W}$ state or $\ket{GHZ}$ state,
as they are two important entanglement classes that are widely used in quantum information applications~\cite{dur2000three}.
And interestingly, 
in the $K>3$ cases, the GCE of the output state from each parallelized permutation test is possibly larger than the values of $\ket{GHZ}$ or $\ket{W}$ states, 
which can be easily tested numerically.
This is because for the GCE of the large system, its maximum value is in general far away from the case of $\ket{GHZ}$ or $\ket{W}$ states, as intensively discussed for $K=2$ in Ref.~\cite{schatzki2024hierarchy}. 
Thus, other different entanglement structures may appear for the controlled system when $K>3$.

\bigskip

\textit{Robustness Analysis.}\textemdash
In this section, we consider the scenario that only noisy states $\ket{\psi'}$ can be prepared such that there is a trace distance $\mathcal{D}(\ket{\psi}, \ket{\psi'}) = \frac{1}{2}\lVert\ket{\psi}\bra{\psi}-\ket{\psi'}\bra{\psi'}\rVert_1 = \epsilon$. 
In this case, the estimated GCE will also have errors accordingly and here we propose an upper-bound for the errors of estimated GCE.

\begin{proposition}\label{prop3}
Suppose there are prime $K$ noisy input copies $\ket{\psi'_1},\ket{\psi'_2},\cdots,\ket{\psi'_K}$ with $\mathcal{D}(\ket{\psi},\ket{\psi'_k})=\epsilon_k$. 
Therefore in general, via the approach shown in Fig. \ref{circfig} and Eq.~\eqref{eqcecompute}, there exists an upper-bound for the estimated GCE error~\footnote{The error definition here is different from the one in~\cite{beckey2021computable} as in this work we focus on the difference between the perfect and noisy scenarios}:

\begin{equation}
\begin{split}
\mathcal{E}=&|\mathcal{C}^{(K)}_{\ket{\psi'_1},\ket{\psi'_2},\cdots,\ket{\psi'_K}}(s)-\mathcal{C}^{(K)}_{\ket{\psi}}(s)| \\
    \leqslant& \frac{2^{|s|}-1}{(K-1)2^{|s|}}\left(\sum_{k=1}^{K}\epsilon_k+\sum_{k<k'}\epsilon_k\epsilon_{k'}+\sum_{k<k'<k''}\epsilon_k\epsilon_{k'}\epsilon_{k''}\right. \\
    & \left.+\cdots+2\prod_{k=1}^K\epsilon_k \right).
    \label{errbound}
\end{split}
\end{equation}
\end{proposition}

For simplicity, we assume each copy of $\ket{\psi}$ is either noisy with $\epsilon$ error, or perfect. 
Illustratively, consider two following examples. 
Firstly, suppose there are $K$ noisy copies, then the GCE error is upper-bounded by:
    \begin{equation}
        \mathcal{E}\leqslant\left(1-\frac{1}{2^{|s|}}\right)\frac{1}{K-1}\left((1+\epsilon)^K+\epsilon^K-1\right).
    \end{equation}
One can easily find the optimal $K$ in either the regime of small $K$ or small $\epsilon$.
Secondly, suppose there is only one noisy copy, the GCE error is upper-bounded by:
    \begin{equation}
        \mathcal{E}\leqslant \left(1 - \frac{1}{2^{|s|}}\right)\frac{1}{K-1} \epsilon.
    \end{equation}
In this case, the upper-bound of $\mathcal{E}$ decreases strongly reciprocally with $K$, but requiring more perfect state copies.

In Fig.~\ref{errorfig}, the numerical simulations for the errors under the two scenarios above are shown. 
Notably, on average, under the same scale of $\mathcal{D}$, the errors $\mathcal{E}$ will decrease under larger number of copies $K$ or with larger state size $n$. 
Moreover, the errors are always below the error bound in Eq.~\eqref{errbound}. 
Note that the error bound in Eq.~\eqref{errbound} is the analytical result and may not be sharp enough. 
It is possible that there also exists a general sharper bound including the system size $n$ as well.

\bigskip

\textit{Properties of GCE.}\textemdash
Our definition of GCE $\mathcal{C}^{(K)}_{\ket{\psi}}(s)$ enjoys some convenient properties, which hold  in the more general setting $K \in {\mathbb R}, K >1$.

\begin{theorem}\label{theorem}
The GCE has the following properties:
\begin{enumerate}
    \item \textit{Pure state entanglement measures: $\mathcal{C}^{(K)}_{\ket{\psi}}(s)$ is non-increasing on average under LOCC.}
    \item \textit{$\mathcal{C}^{(K)}_{\ket{\psi}}(s)=0$ for fully product states $\ket{\psi}=\otimes_{j=1}^{n}\ket{\phi_j}$ and $\forall s\in\mathcal{P}(S)\backslash\{\varnothing\}$.}
    \item \textit{Continuity: For two pure states $\ket{\psi}$ and $\ket{\psi'}$ that satisfy $\mathcal{D}(\ket{\psi},\ket{\psi'})\leqslant\epsilon$, then $\left| \mathcal{C}^{(K)}_{\ket{\psi}}(s)-\mathcal{C}^{(K)}_{\ket{\psi'}}(s) \right|\leqslant \frac{2K}{K-1}\epsilon$.}
    \item \textit{$\mathcal{C}^{(K)}_{\ket{\psi}}(S)=\mathcal{C}^{(K)}_{\ket{\psi}}(S\backslash\{n_0\})$ for any single subsystem label $n_0\in\{1,2,3,...,n\}$.}
\end{enumerate}
\end{theorem}

Theorem~\ref{theorem}.1 shows that $\mathcal{C}^{(K)}_{\ket{\psi}}(s)$ is a well-defined entanglement measure for any $K>1$. 
Also, one should not confuse Theorem~\ref{theorem}.3 with Proposition \ref{prop2}, as Theorem~\ref{theorem}.3 can be regarded as the special case of Proposition \ref{prop2} where the input noisy states are exactly the same. 
The proof of Theorem can be found in Appendix~\ref{theoremproof}.

Furthermore, our numerical studies (c.f. Appendix \ref{NSSSAprop}) provide evidence for the following conjecture: 

\begin{conjecture}\label{conjecture}
The GCE has the following properties:
\begin{enumerate}
    \item \textit{$\mathcal{C}^{(K)}_{\ket{\psi}}(s')\leqslant \mathcal{C}^{(K)}_{\ket{\psi}}(s)$ if $s'\subseteq s$.}
    \item \textit{Subadditivity: $\mathcal{C}^{(K)}_{\ket{\psi}}(s\cup s')\leqslant \mathcal{C}^{(K)}_{\ket{\psi}}(s) + \mathcal{C}^{(K)}_{\ket{\psi}}(s')$ for $s\cap s'=\varnothing$.}
\end{enumerate}
\end{conjecture}

\begin{figure}
    \centering
    \includegraphics[width=1.0\linewidth]{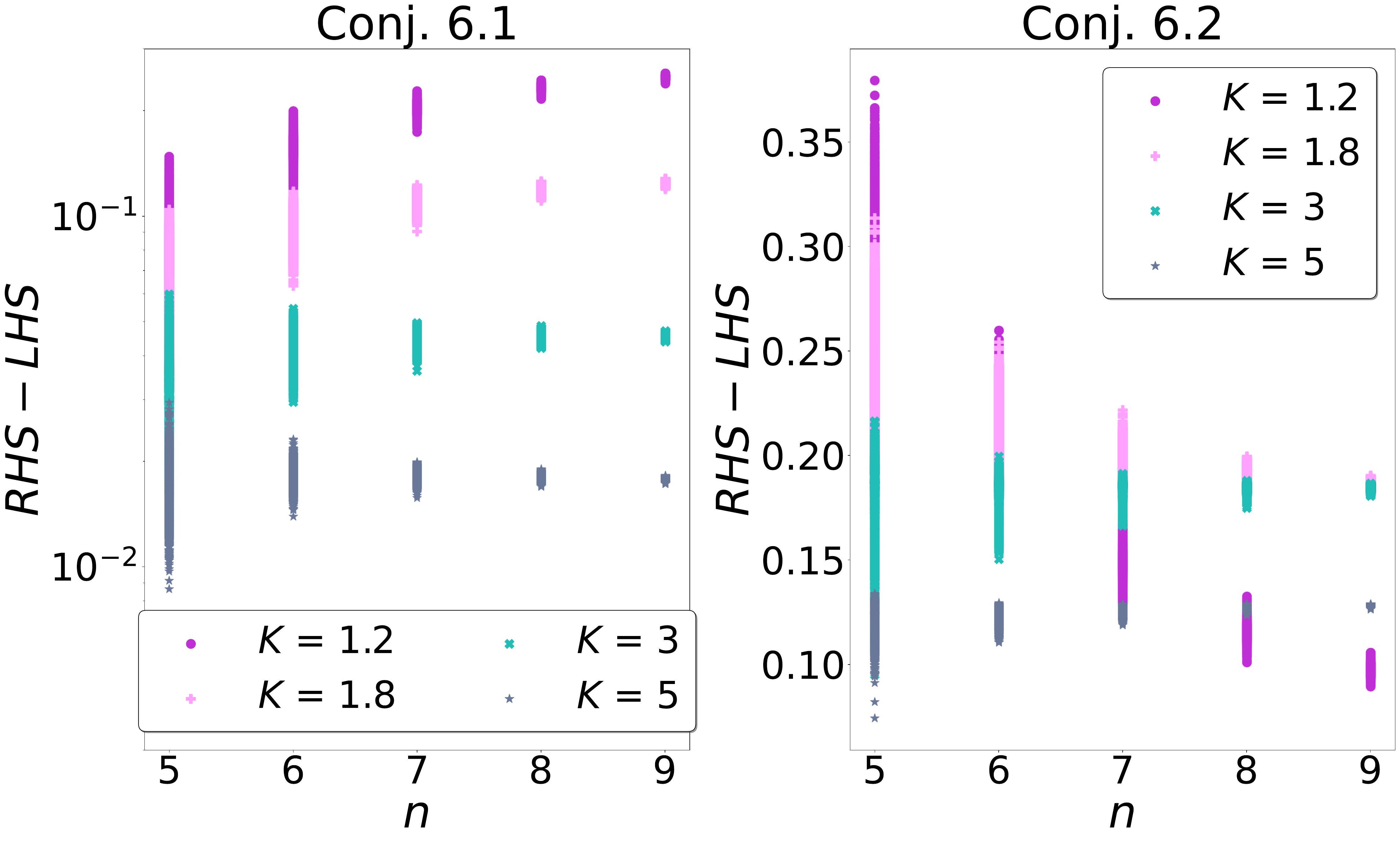}
    \caption{Numerics for Conjectures \ref{conjecture}. The differences between two sides of the inequalities ($RHS-LHS$) are shown for $K=1.2,1.8,3,5$ and $n=5,6,7,8,9$. For Conjecture \ref{conjecture}.1, $(s'=\{1,2,3\})\subseteq (s = \{1,2,3,4\})$. For Conjecture \ref{conjecture}.2, $(s'=\{1,2\})\cap(s=\{3,4\})=\varnothing$. 10000 Haar random states for each set of $\{n,K\}$ are calculated. Numerically, both conjectures hold as every sampled data point is positive. }
    \label{conj}
\end{figure}

Numerical results for Conjecture \ref{conjecture} are shown in Fig.~\ref{conj}. 
Here, Conjecture \ref{conjecture}.1 states that the GCE for any $K>1$ of a given system is always larger than or equal to the GCE of its subsystems, which should hold for a well-defined entanglement measure. 
Conjecture \ref{conjecture}.2 surmises that the GCE for any $K>1$ obeys the subadditivity property, which means that the sum of two separate systems' GCEs should be larger than the GCE of the overall system. 
These two conjectures have been proven to be true for $K=2$~\cite{beckey2021computable}. Based on these arguments we here conjecture that they also hold for $K>1$. 
Moreover, we show that:

\begin{proposition}\label{prop5}
Conjecture~\ref{conjecture}.1 is equivalent to a not-so-strong subadditivity (NSSSA) form of Tsallis entropy: for an $n$-partite (qubit) pure state $\rho = \ket{\psi}\bra{\psi}$ and any of its tri-separation $ABC$ s.t. $B$ contains only one party (qubit), we have:
\begin{equation}
\begin{split}
    \sum_{\alpha_A\in\mathcal{P}(A)}&T_K(\rho_{\alpha_{A}BC})+T_K(\rho_{\alpha_{A}}) \\
    -&T_K(\rho_{\alpha_A B})-T_K(\rho_{\alpha_A C})\leqslant0,
\end{split}
\end{equation}
which is the sum over all possible strong subadditivity (SSA) of Tsallis entropy related to the subsets of $A$.

\end{proposition}

Proposition~\ref{prop5} may be interesting and important in its own right, because SSA of Tsallis entropy does not generally hold~\cite{petz2015inequalities}. 
We refer the reader to Appendix~\ref{discussconjecture} for further discussions.

\bigskip

\begin{figure}
    \centering
    \includegraphics[width=1.0\linewidth]{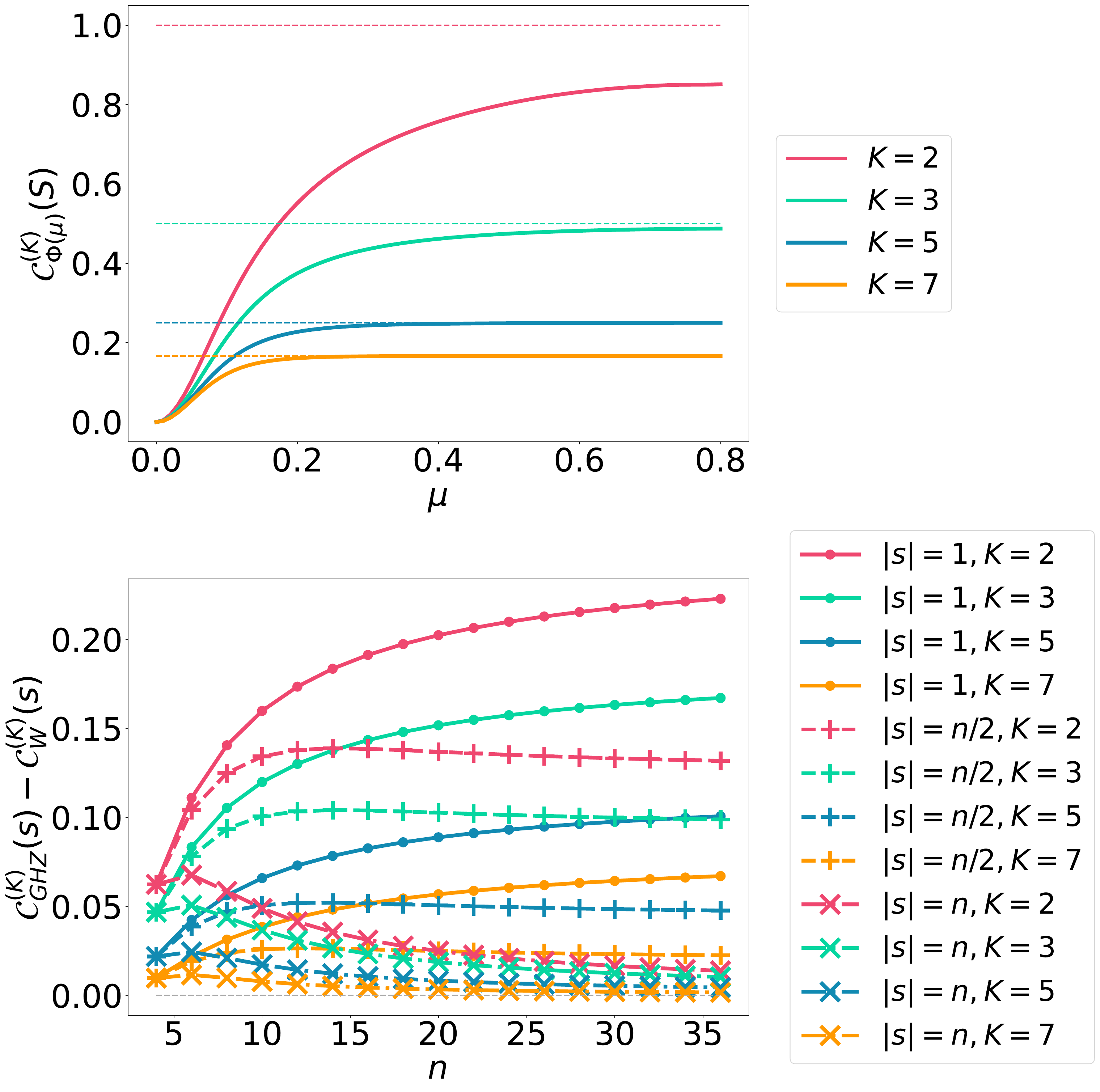}
    \caption{GCE ($s=S$) of the 40-qubit spin-squeezed state $\ket{\Phi(\mu)}$ for different $K$ and interaction strength $\mu$ (upper plot), and the GCE difference $\mathcal{C}^{(K)}_{\ket{GHZ}}(s)-\mathcal{C}^{(K)}_{\ket{W}}(s)$ for different $n$, $|s|$ and $K$ (lower plot). 
    For $\ket{\Phi(\mu)}$, GCE measures more entanglement and goes asymptotically to  $\sim(K-1)^{-1}$ with larger $K$ and $\mu$.   
    For $\ket{GHZ}$ and $\ket{W}$, GCE concludes more entanglement in $\ket{GHZ}$ state than in $\ket{W}$ state as $\mathcal{C}^{(K)}_{\ket{GHZ}}(s)>\mathcal{C}^{(K)}_{\ket{W}}(s)$, and the GCE difference decreases for increasing $K$ and $|s|$.}
    \label{GHZ-W}
\end{figure}

\textit{Examples.}\textemdash
We now calculate GCE for a few selected types of quantum states that are of experimental interest.
Let us first consider spin-squeezed states, $\ket{\Phi(\mu)}$, that can be prepared by evolving a coherent spin state under the one-axis twisting Hamiltonian operator $H_{OAT}=\chi\hat{S}^2_z$ for a time $t$ and parametrized through the interaction strength $\mu=2\chi t$~\cite{kitagawa1993squeezed,guo2023detecting}.
The GCE ($s=S$) of the 40-qubit state $\ket{\Phi(\mu)}$ is shown in Fig.~\ref{GHZ-W} (upper plot). 
Notice that when the interaction strength $\mu$ becomes larger, the GCE predicts more entanglement in the state $\ket{\Phi(\mu)}$. 
Moreover, it goes asymptotically to $\sim(K-1)^{-1}$ for larger $K$, and larger $K$ ends up with faster convergence w.r.t. the interaction strength $\mu$.

Secondly, we consider $W$ and $GHZ$ states. 
The GCE difference between them, $\mathcal{C}^{(K)}_{GHZ}(s)-\mathcal{C}^{(K)}_{W}(s)$, is shown in Fig.~\ref{GHZ-W} (lower plot). 
Note that $\mathcal{C}^{(K)}_{GHZ}(s)>\mathcal{C}^{(K)}_{W}(s)$ for all cases, which illustrates that there is more entanglement in $GHZ$ state than in $W$ state according to GCE measures. 
The GCE difference also decreases with increasing $K$ and $|s|$.
In addition, when $|s|$ is large enough (for example, $|s|\approx n$), increasing $n$ makes the GCE difference go asymptotically to 0. 
More analytical details for these two examples can be found in the Appendix~\ref{examples}. 

\bigskip

\textit{Conclusions and Outlook.}\textemdash
In this work, we proposed a $K$-th order pure-state entanglement measure, \textit{Generalized Concentratable Entanglement}, for arbitrary real-valued $K>1$.
We provided an efficient way for estimating GCE on a quantum computer, through parallelized permutation tests given that the number of state copies, $K$, is prime.
We also showed that each parallelized permutation test can concentrate the entanglement into the $\ket{W}$ state with three state copies.
In addition, we provided both analytical and numerical results for the errors of the estimated GCE when some of the state copies are imperfect.
We demonstrated that these errors become smaller as the number of copies increases, for a constant number of imperfect copies.
Then we proved several mathematical properties of GCE, especially that GCE is a well-defined pure-state entanglement measure as it does not increase under LOCC on average.
Backed up by strong numerical evidence, we also proposed two conjectures that may provide further mathematical insights, not only on GCE, but also of independent interest for the study of quantum Tsallis entropies.
One of them hints at the existence of a weaker form of strong subadditivity of quantum Tsallis entropy (NSSSA), which lies between subadditivity (which holds \cite{audenaert2007subadditivity}) and strong subadditivity (which does not hold in general \cite{petz2015inequalities}), and may serve as an interesting starting point for future investigations.
Finally we provided examples and explicitly calculated GCE for spin-squeezed, $\ket{W}$ and $\ket{GHZ}$ states.

A natural follow-up from our research is to propose an efficient estimation procedure for composite $K$. Furthermore, quantities like $\Tr(\rho^K_{\alpha})$ for any pure state $\rho$ find applicability beyond the GCE context \cite{brandao2005quantifying, shahandeh2017ultrafine, saggio2019experimental, cieslinski2024valid, pal2021bound, ragy2016compatibility}.
Also, it would be interesting to explore the entanglement concentrated from the parallelized permutation test for $K>3$, which remains open~\cite{schatzki2024hierarchy}.
Moreover, deriving a sharper bound for the GCE errors would be useful to study the intrinsic error properties of larger $n$ and $K$.
Finally, a rigorous proof of Conjecture \ref{conjecture} should yield valuable insights in the quest to characterize all entropic inequalities in quantum information science \cite{linden2005new, linden2013structure, cadney2014inequalities, ibinson2007all, morelli2020dimensionally, appel2020monogamy, chitambar2019quantum}.

\bigskip

% ACKNOWLEDGMENTS!!!.
We thank Andreas Winter, Marcus Huber and Jose Carrasco for their valuable insights. 
X.L. thanks Kshiti Sneh Rai and Jinfu Chen for fruitful discussions. 
J.K. gratefully acknowledges support from Dr. Max Rössler, the Walter Haefner Foundation and the ETH Zürich Foundation. 
J.T. acknowledges the support received from the European Union's Horizon Europe research and innovation programme through the ERC StG FINE-TEA-SQUAD (Grant No.~101040729). 
J.T. also acknowledges the support received by the Dutch National Growth Fund (NGF), as part of the Quantum Delta NL programme.
This publication is part of the ‘Quantum Inspire – the Dutch Quantum Computer in the Cloud’ project (with project number [NWA.1292.19.194]) of the NWA research program ‘Research on Routes by Consortia (ORC)’, which is funded by the Netherlands Organization for Scientific Research (NWO).
The views and opinions expressed here are solely those of the authors and do not necessarily reflect those of the funding institutions. Neither of the funding institutions can be held responsible for them.


\bibliography{concentratable_entanglement_1}% Produces the bibliography via BibTeX.

\newpage

\onecolumngrid

\appendix

\section{GCE Circuit for Prime $K$~\label{proofcircuit}}

In the main text we illustrate that the circuit in Fig.~\ref{circfig} can estimate GCE in Eq.~\eqref{defeq} for any prime $K$ and derive some propositions about it. 
We here provide more analytical details.

\subsection{Proof of Proposition \ref{prop1}}

\begin{proof}

We start from analyzing the derangement operator $D$. 
Note that there are in total $K-1$ different $D$. 
Using the two-line notations, we denote the one in Eq.~\eqref{D1} as $D^{(1)}$, along with the rest:

\begin{equation}
D^{(1)}=\begin{pmatrix}
1 & 2 & 3 & \cdots & K \\
2 & 3 & 4 & \cdots & 1
\end{pmatrix}, \ \ 
D^{(2)}=\begin{pmatrix}
1 & 2 & 3 & \cdots & K \\
3 & 4 & 5 & \cdots & 2
\end{pmatrix}, \ \ 
D^{(3)}=\begin{pmatrix}
1 & 2 & 3 & \cdots & K \\
4 & 5 & 6 & \cdots & 3
\end{pmatrix}, \cdots ,
\end{equation}
until,

\begin{equation}
D^{(K-1)}=\begin{pmatrix}
1 & 2 & 3 & \cdots & K \\
K & 1 & 2 & \cdots & K-1
\end{pmatrix}.
\end{equation}

These are all possible cyclic permutations. Also we introduce:

\begin{equation}
D^{(0)}=\begin{pmatrix}
1 & 2 & 3 & \cdots & K \\
1 & 2 & 3 & \cdots & K
\end{pmatrix}=\mathbb{I},
\end{equation}
where no permutations are applied on the qubits at all, forming an identity transformation. 
Moreover, one can easily find that, for any $D^{(z)}$ $(z\in\{0,1,2,\cdots,K-1\})$ and $q\in\mathbb{N}$, we have:

\begin{equation}
\left(D^{(z)}\right)^q=D^{\left(zq \Mod{K}\right)}.
\end{equation}

Therefore, from Eq.~\eqref{Dops}, with the ancilla-control-$D$ applied on the controlled system $\ket{\phi}$, we have:

\begin{equation}
    \sum_{z=0}^{K-1}c_z\ket{z}\otimes\ket{\phi} \rightarrow \sum_{z=0}^{K-1}c_z\ket{z}\otimes D^z \ket{\phi} = \sum_{z=0}^{K-1}c_z\ket{z}\otimes D^{(z)} \ket{\phi}.
\end{equation}

When $K=2$, $D$ becomes the SWAP operator $S$ and we have the simplest SWAP trick:

\begin{equation}
\Tr(S\rho^{\otimes2})=\Tr(\rho^{2}).
\end{equation}
Generalizing the SWAP trick is relatively straightforward, but not in general true for any $D^{(z)}$. 
This is because:\\

\textbf{Lemma 1:} \textit{$\Tr(D^{(z)}\rho^{\otimes K})=\Tr(\rho^{K})$ holds for $\forall z\in\{1,2,\cdots,K-1\}$ and any density matrix $\rho$, if and only if $K$ is prime.}

\begin{proof}

Generally, we can eigendecompose any density matrix $\rho$ in this form:

\begin{equation}
\rho=\sum_{r=1}^{d}p_r\ket{\psi_r}\bra{\psi_r},
\end{equation}
where $d$ is the dimension of the Hilbert space of $\rho$.
Therefore:

\begin{equation}
\Tr(\rho^K)=\Tr\left(\sum_{r_1,\cdots,r_K=1}^{d}p_{r_1}p_{r_2}\cdots p_{r_K}\ket{\psi_{r_1}}\braket{\psi_{r_1}|\psi_{r_2}}\bra{\psi_{r_2}}\cdots\ket{\psi_{r_K}}\bra{\psi_{r_K}}\right)=\sum_{r=1}^{d}p_r^K,
\end{equation}
as the non-zero terms are contributed by the indices $r=r_1=\cdots=r_K$ only.
On the other hand, we have:
\begin{equation}
    \rho^{\otimes K} = \sum_{r_1,\cdots,r_K=1}^{d}p_{r_1}p_{r_2}\cdots p_{r_K}\ket{\psi_{r_1}\psi_{r_2}\cdots\psi_{r_K}}\bra{\psi_{r_1}\psi_{r_2}\cdots\psi_{r_K}}.
\end{equation}
Then, if one of the $D$-s applies on $\rho^{\otimes K}$, the $\psi$-s in the ket part will be permuted and the non-zero terms for $\Tr(D^{(z)}\rho^{\otimes K})$ will be only contributed by satisfying $r_1=r_{\sigma(1)}, r_2=r_{\sigma(2)}, \cdots ,  r_{K}=r_{\sigma(K)}$ simultaneously. 
In order to have $r=r_1=r_2=\cdots=r_K$, the $D$ applied on the $\rho^{\otimes K}$ should have the form of single cyclic permutation.\\

\textbf{Sub-Lemma 1:} \textit{$D^{(z)}$ for $\forall z\in\{1,2,3,\cdots,K-1\}$ can be written as a single cyclic permutation if and only if $K$ is prime. }

\begin{proof}

We divide this proof into two parts. 
Firstly, we show that for $\forall K\in \mathbb{P}$ ($\mathbb{P}$ denotes the prime number set), all $D^{(z)}$ ($z\in\{1,2,3,\cdots,K-1\}$) can be written as a single cyclic permutation (\textit{If} direction).

For any $D^{(z)}$, suppose it is a single cyclic permutation, we have:

\begin{equation}
D^{(z)}=\begin{pmatrix}
1 & 2 & 3 & \cdots & K \\
1+z & 2+z & 3+z & \cdots & K+z
\end{pmatrix} \Mod{K}  =\left(1,1+z,1+2z,\cdots,1+(K-1)z\right) \Mod{K}.
\end{equation}
Note that here in $D^{(z)}$ if an element $x\Mod{K} \equiv 0$ we write $x=K$ instead. 
Now we only need to illustrate that the elements in the single cyclic permutation $D^{(z)}=\left(1,1+z,1+2z,\cdots,1+(K-1)z\right) \Mod{K}$ are completely different from each other. 
By contradictions, if there exists two elements with $z\in\{1,2,\cdots,K-1\}$ s.t.:

\begin{equation}
    1+az \equiv 1+bz \Mod{K} \ \ \ (a,b\in \mathbb{N}, \ K\in \mathbb{P}, \text{ and } 0\leqslant a<b\leqslant K-1),
\end{equation}
then we have:
\begin{equation}
(b-a)z\equiv0 \Mod{K}.
\end{equation}
However, this is not possible because $K\in \mathbb{P}$ and its one and only 2-element factorization is $1\times K$.
Hence, the \textit{If} direction holds.

Secondly, we show that if $K\notin \mathbb{P}$ (i.e., a composite number), then $\exists D^{(z)}$ cannot be represented as a single cyclic permutation (\textit{Only If} direction). 
If $K\notin \mathbb{P}$, we can always find a 2-element factorization for $K$ s.t. $K=lp$ where $p\in \mathbb{P}$ and $l\geqslant2$. 
Consider:

\begin{equation}
\begin{split}
D^{(p)}=&\begin{pmatrix}
1 & 2 & 3 & \cdots & p & 1+p & 2+p & \cdots & K \\
1+p & 2+p & 3+p & \cdots & 2p & 1+2p & 2+2p & \cdots & K+p
\end{pmatrix} \Mod{K}  \\
=&\left(1,1+p,1+2p,\cdots,1+(l-1)p\right)\left(2,2+p,2+2p,\cdots,2+(l-1)p\right)\cdots\left(p,2p,3p,\cdots,lp\right) \Mod{K}.
\end{split}
\end{equation}

The second equality is because $x+lp=x+K\equiv x\Mod{K}$. 
Therefore, there must exist a $D^{(p)}$ for composite number $K$ that cannot be written as a single cyclic permutation. 
Hence, the \textit{Only If} direction holds as well. %(\textbf{Sub-Lemma 1} $\blacksquare$)

\end{proof}

Due to \textbf{Sub-Lemma 1}, we find that only when $K\in\mathbb{P}$, for any $\Tr(D^{(z)}\rho^{\otimes K})$, $r_1=r_{\sigma(1)}, r_2=r_{\sigma(2)}, \cdots ,  r_{K}=r_{\sigma(K)}$ is equivalent to $r=r_1=\cdots=r_K$, thus:
\begin{equation}
    \Tr(D^{(z)}\rho^{\otimes K}) = \Tr(\rho^K) = \sum_{r=1}^{d}p_r^K,
\end{equation}
as we desired. %(\textbf{Lemma 1} $\blacksquare$)

\end{proof}

Also, one can easily find that when $D^{(z)}$ does not permute the entire eigenstate $\ket{\psi}$ but only its reduced systems, we can easily find that: 
\begin{equation}
\Tr(D_{\alpha}^{(z)}\rho^{\otimes K})=\Tr(D_{\alpha}^{(z)}\rho_{\alpha}^{\otimes K})=\Tr(\rho_{\alpha}^{K}),
\end{equation}
holds for $\forall z\in\{1,2,\cdots,K-1\}$, if and only if $K$ is prime. 
Here, $\alpha$ denotes a certain subsystem of $\rho$ and $\rho_{\alpha}$ is the corresponding reduced density matrix. 
$D_{\alpha}^{(z)}$ denotes the permutation ensemble on the subsystem $\alpha$. For example, if $\alpha\in\{1,3,4\}$, then $D_{\alpha}^{(z)}=D_{1}^{(z)}D_{3}^{(z)}D_{4}^{(z)}$, denoting that the permutations act only on the first, third and fourth qubits.

Now, continue the proof for Proposition \ref{prop1}. 
As for each parallelized permutation test, suppose the controlled state is $\ket{\Psi_j}$, then the state becomes (before measurement on the ancilla):

\begin{equation}
\frac{1}{K}\sum_{z_j=0}^{K-1}\ket{z_j}\sum_{m=0}^{K-1}\bar{\omega}^{z_jm}D_j^{(m)}\ket{\Psi_j},
\end{equation}
where $\bar{\omega}=\omega^{*}=e^{-2\pi i / K}$. 
After measurement that clicks on $\ket{z_j}$, the corresponding operation $\sum_{m=0}^{K-1}\bar{\omega}^{z_jm}D_j^{(m)}$ will be applied on $\ket{\Psi_j}$ finally. 
We consider this process as a quantum channel with Kraus operators:

\begin{equation}
\mathbf{K}_{z_j}=\frac{1}{K}\sum_{m=0}^{K-1}\bar{\omega}^{z_jm}D_j^{(m)}.
\end{equation}

Then, consider the circuit in Fig.~\ref{circfig} as a whole, the overall Kraus operators can be written as:

\begin{equation}
\mathbf{K}_{\mathbf{z}}=\mathbf{K}_{z_1}\mathbf{K}_{z_2}\cdots \mathbf{K}_{z_n}=\frac{1}{K^n}\prod_{j=1}^{n}\left( \sum_{m=0}^{K-1}\bar{\omega}^{z_j m}D_{j}^{(m)} \right).
\end{equation}

Therefore, the probability of getting the measurement result with digit string $\mathbf{z}$ is:

\begin{equation}
p(\mathbf{z}) = \bra{\Psi}\mathbf{K}_{\mathbf{z}}^{\dagger}\mathbf{K}_{\mathbf{z}}\ket{\Psi},
\end{equation}

where $\ket{\Psi} = \ket{\psi}^{\otimes K}$. 
One can easily find that $\mathbf{K}_{\mathbf{z}}=\mathbf{K}_{\mathbf{z}}^{\dagger}\mathbf{K}_{\mathbf{z}}$, therefore:

\begin{equation}
p(\mathbf{z}) = \bra{\Psi}\mathbf{K}_{\mathbf{z}}\ket{\Psi}=\bra{\Psi}\frac{1}{K^n}\prod_{j=1}^{n}\left( \sum_{m=0}^{K-1}\bar{\omega}^{z_j m}D_{j}^{(m)} \right)\ket{\Psi}.
\end{equation}

Suppose we want to measure the entanglement on qubit labels $\alpha=\{\alpha_1,\alpha_2,\cdots,\alpha_l\}$ and denote the rest of the qubit labels as the complementary $\bar{\alpha}=\{\bar{\alpha}_1,\bar{\alpha}_2,\cdots,\bar{\alpha}_{\bar{l}}\}=S \backslash \alpha$.
Certainly, we have $l+\bar{l}=n$. 

Now, consider the quantity:

\begin{equation}
\begin{split}
&\sum_{\sum_{x=1}^{l}z_{\alpha_x}\equiv0\Mod{K}}p(\boldsymbol{z}) \\
=& \frac{1}{K^n}\bra{\Psi}\left[\sum_{z_{\bar{\alpha}_1},\cdots,z_{\bar{\alpha}_{\bar{l}}}=0}^{K-1}\prod_{\bar{u}=\bar{\alpha}_1}^{\bar{\alpha}_{\bar{l}}}\left( \sum_{m=0}^{K-1}\bar{\omega}^{z_{\bar{u}} m}D_{\bar{u}}^{(m)} \right)\right]\left[\sum_{\sum_{x=1}^{l}z_{\alpha_x}\equiv0\Mod{K}}\prod_{u=\alpha_1}^{\alpha_l}\left( \sum_{m=0}^{K-1}\bar{\omega}^{z_u m}D_{u}^{(m)} \right)\right]\ket{\Psi} \\ 
=&\frac{1}{K^n}\bra{\Psi}\mathbf{F}\mathbf{G}\ket{\Psi}.
\end{split}
\end{equation}

Here we separate the inner filling of the bra-ket sandwich into two parts: $\mathbf{F}$ and $\mathbf{G}$, for non-tested registers ($\in\bar{\alpha}$) and the tested registers ($\in\alpha$), respectively. 
We start from calculating $\mathbf{F}$:

\begin{equation}
\begin{split}
\mathbf{F}=&\sum_{z_{\bar{\alpha}_1},\cdots,z_{\bar{\alpha}_{\bar{l}}}=0}^{K-1}\prod_{\bar{u}=\bar{\alpha}_1}^{\bar{\alpha}_{\bar{l}}}\left( \sum_{m=0}^{K-1}\bar{\omega}^{z_{\bar{u}} m}D_{\bar{u}}^{(m)} \right)    \\
=&\sum_{z_{\bar{\alpha}_1},\cdots,z_{\bar{\alpha}_{\bar{l}}}=0}^{K-1}\left( \sum_{m=0}^{K-1}\bar{\omega}^{z_{\bar{\alpha}_1} m}D_{\bar{\alpha}_1}^{(m)} \right)\left( \sum_{m=0}^{K-1}\bar{\omega}^{z_{\bar{\alpha}_2} m}D_{\bar{\alpha}_2}^{(m)} \right)\cdots\left( \sum_{m=0}^{K-1}\bar{\omega}^{z_{\bar{\alpha}_{\bar{l}}} m}D_{\bar{\alpha}_{\bar{l}}}^{(m)} \right) \\
=&\left[ \sum_{m=0}^{K-1}D_{\bar{\alpha}_1}^{(m)}\left( \sum_{z_{\bar{\alpha}_1}=0}^{K-1}\bar{\omega}^{z_{\bar{\alpha}_1} m} \right) \right]\left[ \sum_{m=0}^{K-1}D_{\bar{\alpha}_2}^{(m)}\left( \sum_{z_{\bar{\alpha}_2}=0}^{K-1}\bar{\omega}^{z_{\bar{\alpha}_2} m} \right) \right]\cdots\left[ \sum_{m=0}^{K-1}D_{\bar{\alpha}_{\bar{l}}}^{(m)}\left( \sum_{z_{\bar{\alpha}_{\bar{l}}}=0}^{K-1}\bar{\omega}^{z_{\bar{\alpha}_{\bar{l}}} m} \right) \right].
\end{split}
\end{equation}

Since
\begin{equation}
\sum_{z=0}^{K-1}\bar{\omega}^{mz}=\begin{cases} 
  K, & \text{if } m=0 \\
  0, & \text{if } m\in\{1,2,\cdots,K-1\} 
\end{cases},
\end{equation}
we have
\begin{equation}
\mathbf{F}=K^{\bar{l}} D_{\bar{\alpha}_1}^{(0)}D_{\bar{\alpha}_2}^{(0)}\cdots D_{\bar{\alpha}_{\bar{l}}}^{(0)} = K^{\bar{l}}\mathbb{I}.
\end{equation}

Secondly, for $\mathbf{G}$:
\begin{equation}
\begin{split}
\mathbf{G}=&\sum_{\sum_{x=1}^{l}z_{\alpha_x}=0\mod K}\prod_{u=\alpha_1}^{\alpha_l}\left( \sum_{m=0}^{K-1}\bar{\omega}^{z_u m}D_{u}^{(m)} \right) \\
=&\sum_{\sum_{x=1}^{l}z_{\alpha_x}=0\mod K}\left( \sum_{m=0}^{K-1}\bar{\omega}^{z_{\alpha_1} m}D_{\alpha_1}^{(m)} \right)\left( \sum_{m=0}^{K-1}\bar{\omega}^{z_{\alpha_2} m}D_{\alpha_2}^{(m)} \right)\cdots\left( \sum_{m=0}^{K-1}\bar{\omega}^{z_{\alpha_l} m}D_{\alpha_l}^{(m)} \right).
\end{split}
\end{equation}

Here, we cannot freely choose all $z_{\alpha_x}$ since the summation of $z_{\alpha_x}$ is bounded by the condition of $\sum_{x=1}^{l}z_{\alpha_x}\equiv0\Mod{K}$. 
However, since $z_{\alpha_l}\equiv-\sum_{x=1}^{l-1}z_{\alpha_x}\Mod{K}$, once we freely choose $z_{\alpha_1},z_{\alpha_2},\cdots,z_{\alpha_{l-1}}\in\{0,1,\cdots,K-1\}$, $z_{\alpha_l}$ will be one and only determined. 
We now denote $z_{\alpha_l}=aK-\sum_{x=1}^{l-1}z_{\alpha_x}$ $(a\in\mathbb{Z})$, then:

\begin{equation}
\begin{split}
\mathbf{G}=&\sum_{z_{\alpha_1},z_{\alpha_2},\cdots,z_{\alpha_{l-1}}=0}^{K-1}\sum_{m_1,m_2,\cdots,m_l=0}^{K-1}\bar{\omega}^{(m_1-m_l)z_{\alpha_1}+(m_2-m_l)z_{\alpha_2}+\cdots+(m_{l-1}-m_l)z_{\alpha_{l-1}}}D_{\alpha_1}^{(m_1)}D_{\alpha_2}^{(m_2)}\cdots D_{\alpha_l}^{(m_l)} \\
=&\sum_{m_1,m_2,\cdots,m_l=0}^{K-1}D_{\alpha_1}^{(m_1)}D_{\alpha_2}^{(m_2)}\cdots D_{\alpha_l}^{(m_l)}\left(\sum_{z_{\alpha_1}=0}^{K-1}\bar{\omega}^{(m_1-m_l)z_{\alpha_1}}\right)\left(\sum_{z_{\alpha_2}=0}^{K-1}\bar{\omega}^{(m_2-m_l)z_{\alpha_2}}\right)\cdots\left(\sum_{z_{\alpha_{l-1}}=0}^{K-1}\bar{\omega}^{(m_{l-1}-m_l)z_{\alpha_{l-1}}}\right).
\end{split}
\end{equation}

The terms are non-trivial if and only if $m=m_1=m_2=\cdots=m_l\in\{0,1,2,\cdots,K-1\}$. 
Therefore:
\begin{equation}
\mathbf{G} = K^{l-1}\sum_{m=0}^{K-1}D_{\alpha_1}^{(m)}D_{\alpha_2}^{(m)}\cdots D_{\alpha_l}^{(m)}.
\end{equation}

Therefore, when $K\in\mathbb{P}$, we have:
\begin{equation}
\begin{split}
\sum_{\sum_{x=1}^{l}z_{\alpha_x}\equiv0\Mod{K}}p(\mathbf{z})=&\frac{1}{K^n}\bra{\Psi}\mathbf{F}\mathbf{G}\ket{\Psi} =\frac{K^{l+\bar{l}-1}}{K^n}\Tr\left(\sum_{m=0}^{K-1}D_{\alpha_1}^{(m)}D_{\alpha_2}^{(m)}\cdots D_{\alpha_l}^{(m)}\ket{\Psi}\bra{\Psi}\right) \\
=&\frac{1}{K}\Tr\left(\left(\mathbb{I}+\sum_{m=1}^{K-1}D^{(m)}_{\alpha}\right)\rho^{\otimes K}\right) =\frac{1 + (K-1)\Tr(\rho_{\alpha}^{K})}{K}.
\end{split}
\end{equation}

Therefore:
\begin{equation}
\Tr(\rho_{\alpha}^{K}) = \frac{1}{K-1} \left(K \sum_{\sum_{x=1}^{l}z_{\alpha_x}\equiv0\Mod{K}}p(\mathbf{z}) - 1\right) = \frac{1}{K-1}\left( K\sum_{\sum_{x\in\alpha }z_x\equiv0 \Mod{K}}p(\mathbf{z})-1 \right).
\end{equation}

Now, consider another quantity:

\begin{equation}
\begin{split}
&\sum_{\sum_{x=1}^{l}z_{\alpha_x}\not\equiv0\Mod{K}}p(\boldsymbol{z}) \\
=& \frac{1}{K^n}\bra{\Psi}\left[\sum_{z_{\bar{\alpha}_1},\cdots,z_{\bar{\alpha}_{\bar{l}}}=0}^{K-1}\prod_{\bar{u}=\bar{\alpha}_1}^{\bar{\alpha}_{\bar{l}}}\left( \sum_{m=0}^{K-1}\bar{\omega}^{z_{\bar{u}} m}D_{\bar{u}}^{(m)} \right)\right]\left[\sum_{\sum_{x=1}^{l}z_{\alpha_x}\not\equiv0\Mod{K}}\prod_{u=\alpha_1}^{\alpha_l}\left( \sum_{m=0}^{K-1}\bar{\omega}^{z_u m}D_{u}^{(m)} \right)\right]\ket{\Psi} \\ 
=&\frac{1}{K^n}\bra{\Psi}\mathbf{F}\mathbf{G'}\ket{\Psi},
\end{split}
\end{equation}

where $\mathbf{F}$ did not change but $\mathbf{G}$ changed into $\mathbf{G'}$. 
Suppose  $\sum_{x=1}^{l}z_{\alpha_x}\equiv t\Mod{K}$ where $t\neq 0$. 
Then,  $z_{\alpha_l}\equiv t-\sum_{x=1}^{l-1}z_{\alpha_x}\Mod{K}$.
Similarly, once we freely choose $z_{\alpha_1},z_{\alpha_2},\cdots,z_{\alpha_{l-1}}\in\{0,1,\cdots,K-1\}$, $z_{\alpha_l}$ will be one and only determined. 
Similar to previous method, we let $z_{\alpha_l}=aK+t-\sum_{x=1}^{l-1}z_{\alpha_x}$ $(a\in\mathbb{Z})$, then:

\begin{equation}
\begin{split}
\mathbf{G'}=&\sum_{z_{\alpha_1},\cdots,z_{\alpha_{l-1}}=0}^{K-1}\sum_{m_1,\cdots,m_l=0}^{K-1}\bar{\omega}^{tm_l}\bar{\omega}^{(m_1-m_l)z_{\alpha_1}+(m_2-m_l)z_{\alpha_2}+\cdots+(m_{l-1}-m_l)z_{\alpha_{l-1}}}D_{\alpha_1}^{(m_1)}D_{\alpha_2}^{(m_2)}\cdots D_{\alpha_l}^{(m_l)} \\
=&\sum_{m_1,\cdots,m_l=0}^{K-1}\bar{\omega}^{tm_l}D_{\alpha_1}^{(m_1)}D_{\alpha_2}^{(m_2)}\cdots D_{\alpha_l}^{(m_l)}\left(\sum_{z_{\alpha_1}=0}^{K-1}\bar{\omega}^{(m_1-m_l)z_{\alpha_1}}\right)\left(\sum_{z_{\alpha_2}=0}^{K-1}\bar{\omega}^{(m_2-m_l)z_{\alpha_2}}\right)\cdots\left(\sum_{z_{\alpha_{l-1}}=0}^{K-1}\bar{\omega}^{(m_{l-1}-m_l)z_{\alpha_{l-1}}}\right) \\
=&K^{l-1}\sum_{m=0}^{K-1}\bar{\omega}^{mt} D_{\alpha_1}^{(m)}D_{\alpha_2}^{(m)}\cdots D_{\alpha_l}^{(m)}.
\end{split}
\end{equation}

Then:

\begin{equation}
\begin{split}
\sum_{\sum_{x=1}^{l}z_{\alpha_x}\not\equiv 0\Mod{K}}p(\boldsymbol{z})=&\frac{1}{K^n}\bra{\Psi}FG'\ket{\Psi} =\frac{1}{K}\left(1+\sum_{m=1}^{K-1}\bar{\omega}^{mt}\Tr(\rho_{\alpha}^K)\right) \\
=&\frac{1}{K}\left(1+\Tr(\rho_{\alpha}^K)\sum_{m=0}^{K-1}\bar{\omega}^{mt}-\Tr(\rho_{\alpha}^K)\right) =\frac{1-\Tr(\rho_{\alpha}^K)}{K}.
\end{split}
\end{equation}

This also shows that $\sum_{\sum_{x=1}^{l}z_{\alpha_x}\not\equiv 0\Mod{K}}p(\boldsymbol{z})$ is the same for $\forall t\neq 0$ that satisfies $\sum_{x=1}^{l}z_{\alpha_x}\equiv t \Mod{K}$. 
Then:

\begin{equation}
\Tr(\rho_{\alpha}^K)=1-K\sum_{\sum_{x=1}^{l}z_{\alpha_x}\not\equiv 0\Mod {K}}p(\boldsymbol{z}) = 1-K \sum_{\sum_{x\in\alpha}z_x\not\equiv 0 \Mod{K}} p(\mathbf{z}).
\end{equation}

With simple algebra, we then have:

\begin{equation}
    \mathcal{C}^{(K)}_{\ket{\psi}}(s) =\frac{K}{2^{|s|}(K-1)^2}\sum_{\alpha\in\mathcal{P}(s)}\left(1-\sum_{\sum_{x\in\alpha}z_x\equiv 0 \Mod{K}}p(\mathbf{z})\right) =\frac{K}{2^{|s|}(K-1)}\sum_{\alpha\in\mathcal{P}(s)}\sum_{\sum_{x\in\alpha}z_x\not\equiv 0 \Mod{K}}p(\mathbf{z}).
\end{equation}

%(\textbf{Proposition \ref{prop1}} $\blacksquare$).

\end{proof}

One important comment is that one may consider that in order to make this work for composite $K$, different power of $D$ should be controlled by the ancilla instead of $D^{(z)}$ always. 
However, in this way $\mathbf{K}_{\mathbf{z}}\neq\mathbf{K}_{\mathbf{z}}^{\dagger}\mathbf{K}_{\mathbf{z}}$ in general and the proof above does not hold anymore.

\subsection{Proof of Corollary \ref{corollary}}

\begin{proof}

Suppose $\alpha$ covers all qubit labels, i.e. $\alpha=S$. 
Then we have (remember $\rho$ is pure):

\begin{equation}
\sum_{z'_1+z'_2+\cdots+z'_n\equiv0\Mod{K}}p(\mathbf{z'})=\frac{1+(K-1)\Tr(\rho^K)}{K}=1.
\end{equation}

Then:

\begin{equation}
\sum_{z'_1+z'_2+\cdots+z'_n\not\equiv0\Mod{K}}p(\mathbf{z'})=1-\sum_{z'_1+z'_2+\cdots+z'_n\equiv0\Mod{K}}p(\mathbf{z'})=0.
\end{equation}

As $p(\mathbf{z'})$ is always positive, then $p(\mathbf{z'})=0$ for any $h(\mathbf{z'})\not\equiv 0 \Mod{K}$. %(\textbf{Corollary \ref{corollary}} $\blacksquare$).

\end{proof}

\subsection{Proof of Proposition \ref{prop2}}

\begin{proof}

For each parallelized permutation test, as shown in Eq.~\eqref{prop2}, the inputs $\mathbf{1}_j,\mathbf{2}_j,\mathbf{3}_j$ are independent from each other at the beginning as they are from different copies of state $\ket{\psi}$. 
In general, we can consider the state input as:

\begin{equation}
    \mathbf{1}_j\otimes\mathbf{2}_j\otimes\mathbf{3}_j = (a_1\ket{0}+b_1\ket{1})\otimes(a_2\ket{0}+b_2\ket{1})\otimes(a_3\ket{0}+b_3\ket{1}).
\end{equation}

Firstly, we suppose the top ancilla clicks at $\ket{1}$. 
Then, the operator applied on the input state is $\mathbb{I}+\bar{\omega}D^{(1)}_{j}+\bar{\omega}^2 D^{(2)}_{j}$ and the input becomes (unnormalized):

\begin{equation}
\begin{split}
& a_1a_2a_3\left(\frac{b_1}{a_1}+\bar{\omega}\frac{b_2}{a_2}+\bar{\omega}^2\frac{b_3}{a_3}\right)(\ket{100}+\omega\ket{010}+\omega^2\ket{001})+b_1b_2b_3\left(\frac{a_1}{b_1}+\bar{\omega}\frac{a_2}{b_2}+\bar{\omega}^2\frac{a_3}{b_3}\right)(\ket{011}+\omega\ket{101}+\omega^2\ket{110}) \\
    =& \mathbf{M} (\ket{100}+\omega\ket{010}+\omega^2\ket{001}) + \mathbf{N} (\ket{011}+\omega\ket{101}+\omega^2\ket{110}).
\end{split}
\end{equation}

As $K=3$, then $\omega=e^{2\pi i /3}$ and $\bar{\omega}=e^{-2\pi i /3}$. 
Now we apply the local unitaries on each party. 
The general single-qubit gate $U$ is denoted as:

\begin{equation}
U(\theta,\phi, \lambda) = \begin{bmatrix}
                            \cos{\frac{\theta}{2}} & -e^{i\lambda}\sin{\frac{\theta}{2}}\\
                            e^{i\phi}\sin{\frac{\theta}{2}} & e^{i(\phi+\lambda)}\cos{\frac{\theta}{2}}
                            \end{bmatrix}.
\end{equation}

Suppose we fix $\lambda$ and $\theta$ for all three different unitaries, but only change the angle $\phi$, i.e. $U_1 = U(\theta, \phi_1, \lambda)$, $U_2 = U(\theta, \phi_2, \lambda)$ and $U_3 = U(\theta, \phi_3, \lambda)$. 
Then the state becomes:

\begin{equation}
    \begin{split}
        &\left(\mathbf{M}\omega^2 e^{i(\phi_3+\lambda)} \cos{\frac{\theta}{2}} + \mathbf{N}\omega^2 e^{i(\phi_3+2\lambda)} \sin{\frac{\theta}{2}}\right)\ket{001}+ \\
        &\left(\mathbf{M}\omega   e^{i(\phi_2+\lambda)} \cos{\frac{\theta}{2}} + \mathbf{N}\omega   e^{i(\phi_2+2\lambda)} \sin{\frac{\theta}{2}}\right)\ket{010}+ \\
        &\left(\mathbf{M}               e^{i(\phi_1+\lambda)} \cos{\frac{\theta}{2}} + \mathbf{N}               e^{i(\phi_1+2\lambda)} \sin{\frac{\theta}{2}}\right)\ket{100}+ \\
        &\left(-\mathbf{M}\omega^2 e^{i(\phi_1+\phi_2+\lambda)} \sin{\frac{\theta}{2}} + \mathbf{N}\omega^2 e^{i(\phi_1+\phi_2+2\lambda)} \cos{\frac{\theta}{2}}\right)\ket{110}+ \\
        &\left(-\mathbf{M}\omega   e^{i(\phi_1+\phi_3+\lambda)} \sin{\frac{\theta}{2}} + \mathbf{N}\omega   e^{i(\phi_1+\phi_3+2\lambda)} \cos{\frac{\theta}{2}}\right)\ket{101}+ \\
        &\left(-\mathbf{M}               e^{i(\phi_2+\phi_3+\lambda)} \sin{\frac{\theta}{2}} + \mathbf{N}               e^{i(\phi_2+\phi_3+2\lambda)} \cos{\frac{\theta}{2}}\right)\ket{011} \\
       =&e^{i\lambda}\left(\mathbf{M}\cos{\frac{\theta}{2}}+\mathbf{N}e^{i\lambda}\sin{\frac{\theta}{2}}\right) \left(\omega^2e^{i\phi_3}\ket{001} + \omega e^{i\phi_2}\ket{010} + e^{i\phi_1}\ket{100}\right) + \\
        &e^{i\lambda}\left(\mathbf{N}e^{i\lambda}\cos{\frac{\theta}{2}} - \mathbf{M}\sin{\frac{\theta}{2}}\right) \left(\omega^2e^{i(\phi_1+\phi_2)}\ket{110} + \omega e^{i(\phi_1+\phi_3)}\ket{101} + e^{i(\phi_2+\phi_3)}\ket{011}\right).
    \end{split}
\end{equation}

In order to get $W$ state $\frac{1}{\sqrt{3}}(\ket{001}+\ket{010}+\ket{100})$ up to global phases, we can let $e^{i\phi_1}= 1$, $e^{i\phi_2}= \omega^2$, $e^{i\phi_3}= \omega$ first. 
Then, we let:

\begin{equation}
\mathbf{N}e^{i\lambda}\cos{\frac{\theta}{2}} - \mathbf{M}\sin{\frac{\theta}{2}} = 0 \rightarrow \tan{\frac{\theta}{2}} = \frac{\mathbf{N}}{\mathbf{M}}e^{i\lambda}.
\end{equation}

We can let the $\lambda$ that makes $\frac{\mathbf{N}}{\mathbf{M}}e^{i\lambda}$ real, then find the corresponding $\theta$. 
In this way, we illustrate that the output state on ancilla $\ket{1}$ can be transformed to $W$ state via local unitaries. 
Also, for ancilla $\ket{2}$, the proof is similar since due to symmetries, we only need to swap the terms that include $\omega$ and $\omega^2$ and everything else goes completely the same. %(\textbf{Proposition \ref{prop2}} $\blacksquare$). 

\end{proof}

\subsection{Proof of Proposition \ref{prop3}}

\begin{proof}

Consider the scenario where the input state copies are not exactly the same. 
Suppose they are $\ket{\psi'_1},\ket{\psi'_2},\cdots,\ket{\psi'_K}$ with corresponding pure state density matrices $\rho'_1, \rho'_2, \cdots, \rho'_K$, respectively. 
Then, from Proposition \ref{prop1}, one can find that:

\begin{equation}
\begin{split}
\sum_{\sum_{x=1}^{l}z_{\alpha_x}\equiv0\Mod{K}}p(\mathbf{z}) =& \frac{1}{K}\Tr\left(\sum_{m=0}^{K-1}D_{\alpha_1}^{(m)}D_{\alpha_2}^{(m)}\cdots D_{\alpha_l}^{(m)}\ket{\Psi}\bra{\Psi}\right) \\
=& \frac{1}{K} \sum_{m=0}^{K-1}\Tr\left(D_{\alpha_1}^{(m)}D_{\alpha_2}^{(m)}\cdots D_{\alpha_l}^{(m)} \rho'_{1} \otimes \rho'_{2} \otimes \cdots \otimes \rho'_{K} \right) \\
=& \frac{1}{K} \left( 1 + \sum_{k=1}^{K-1}\Tr\left(  \rho'_{1,\alpha}\rho'_{(1+k \Mod{K}),\alpha}\cdots\rho'_{(1+(K-1)k \Mod{K}),\alpha}  \right)   \right). \\
\end{split}
\end{equation}

In this proof we denote $aK \Mod{K}\equiv K$ for $\forall a\in\mathbb{Z}$. 
Then, the result $\mathcal{C}^{(K)}_{\ket{\psi'_1},\ket{\psi'_2},\cdots,\ket{\psi'_K}}(s)$ becomes:

\begin{equation}
\begin{split}
    &\mathcal{C}^{(K)}_{\ket{\psi'_1},\ket{\psi'_2},\cdots,\ket{\psi'_K}}(s) \\
    =& \frac{1}{K-1}\left( 1-\frac{1}{2^{|s|}}\sum_{\alpha\in P(s)}\left(  \frac{1}{K-1}\sum_{k=1}^{K-1}\Tr\left(  \rho'_{1,\alpha}\rho'_{(1+k \Mod{K}),\alpha}\cdots\rho'_{(1+(K-1)k \Mod{K}),\alpha}  \right) \right) \right).
\end{split}
\end{equation}

Since $\mathcal{C}^{(K)}_{\ket{\psi}}(s) = \frac{1}{K-1}\left(1 - \frac{1}{2^{|s|}}\sum_{\alpha\in\mathcal{P}(s)}\Tr(\rho_{\alpha}^K)\right)$, then the error:

\begin{equation}
\begin{split}
    \mathcal{E} =& \left| \mathcal{C}^{(K)}_{\ket{\psi'_1},\ket{\psi'_2},\cdots,\ket{\psi'_K}}(s)-\mathcal{C}^{(K)}_{\ket{\psi}}(s) \right| \\
                =& \frac{1}{(K-1)2^{|s|}}\left| \sum_{\alpha\in\mathcal{P}(s)\backslash \varnothing}\left( \Tr(\rho_{\alpha}^K)-\frac{1}{K-1}\sum_{k=1}^{K-1}\Tr\left(  \rho'_{1,\alpha}\rho'_{(1+k \Mod{K}),\alpha}\cdots\rho'_{(1+(K-1)k \Mod{K}),\alpha} \right) \right) \right| \\
                \leqslant & \frac{1}{(K-1)2^{|s|}} \sum_{\alpha\in\mathcal{P}(s)\backslash \varnothing} \left|\Tr(\rho_{\alpha}^K)-\frac{1}{K-1}\sum_{k=1}^{K-1}\Tr\left(  \rho'_{1,\alpha}\rho'_{(1+k \Mod{K}),\alpha}\cdots\rho'_{(1+(K-1)k \Mod{K}),\alpha} \right)\right|.
\end{split}
\end{equation}

Then the key step is to find the upper-bound of the following quantity:

\begin{equation}
    \left|\Tr(\rho_{\alpha}^K)-\frac{1}{K-1}\sum_{k=1}^{K-1}\Tr\left(  \rho'_{1,\alpha}\rho'_{(1+k \Mod{K}),\alpha}\cdots\rho'_{(1+(K-1)k \Mod{K}),\alpha} \right)\right|.
\end{equation}

As $\mathcal{D}(\ket{\psi},\ket{\psi'_{k}})=\mathcal{D}(\rho,\rho'_{k})=\epsilon_k$, we then have:

\begin{equation}
\mathcal{D}(\rho_{\alpha},\rho'_{k,\alpha})=\frac{1}{2}\lVert \rho_{\alpha}-\rho'_{k,\alpha} \rVert_1=\epsilon_{k,\alpha}\leqslant\epsilon_k   .
\end{equation}

Suppose we set $\rho'_{k,\alpha}=\rho_{\alpha}+2\epsilon_{k,\alpha}W_{k,\alpha}$, then the perturbation $W_{k,\alpha}$ has the following properties:

\begin{enumerate}
    \item $W_{k,\alpha}$ is a Hermitian, i.e., $W_{k,\alpha}=W_{k,\alpha}^{\dagger}$.
    \item $\Tr(W_{k,\alpha})=0$.
    \item $\lVert W_{k,\alpha} \rVert_1=1$.
    \item $\lVert W_{k,\alpha} \rVert_{\infty}\leqslant\frac{1}{2}$.  This is because due to the previous two properties, we have the following equality for the eigenvalues of the perturbation $W_{k,\alpha}$:
    
    \begin{equation}
        \sum_{i^{+}}\lambda^{+}_{i^{+}}+\sum_{i^{-}}\lambda^{-}_{i^{-}} = 0, \\     
    \end{equation}

    \begin{equation}
        \sum_{i^{+}}\left|\lambda^{+}_{i^{+}}\right|+\sum_{i^{-}}\left|\lambda^{-}_{i^{-}}\right| = 1 ,  
    \end{equation}
    where $\lambda^{+}_{i^{+}}$ and $\lambda^{-}_{i^{-}}$ are the positive and negative eigenvalues of $W_{k,\alpha}$, respectively. Therefore:
    \begin{equation}
    \sum_{i^{+}}\left|\lambda^{+}_{i^{+}}\right| = \sum_{i^{+}}\lambda^{+}_{i^{+}}=-\sum_{i^{-}}\lambda^{-}_{i^{-}} = \sum_{i^{-}}\left|\lambda^{-}_{i^{-}}\right|,
    \end{equation}
    and also:
    \begin{equation}
    \lVert W_{k,\alpha}\rVert_{\infty}=\max_{i^{+}}{\lambda_{i^{+}}}\leqslant\sum_{i^{+}}\lambda^{+}_{i^{+}}=\frac{1}{2}.
    \end{equation}  
\end{enumerate}

For the upper-bound calculations, we also use several other important inequalities as follows:

\begin{enumerate}
    \item $\lVert\rho^k\rVert_{1}\leqslant 1$ and $\lVert\rho^k\rVert_{\infty}\leqslant 1$ for any density matrix $\rho$.
    \item $|\Tr(A^{\dagger}B)|\leqslant \lVert A\rVert_p \lVert B\rVert_q$ where $p,q>0$ and $p^{-1}+q^{-1}=1$ for any $n\times n$ matrices $A$ and $B$ (Hölder's inequality for Schatten norms).
    \item $\lVert AB\rVert_{\infty}\leqslant\lVert A\rVert_{\infty} \lVert B\rVert_{\infty}$ for any $n\times n$ matrices $A$ and $B$.
\end{enumerate}

Now we start finding the upper-bound. 
We first illustrate the simplest case when $K=2$ as this is the smallest and the only even prime number. 
Then,

\begin{equation}
\begin{split}
    &\left| \Tr(\rho_{\alpha}^2) - \Tr(\rho'_{1,\alpha}\rho'_{2,\alpha}) \right| \\
  =&\left| 2\epsilon_{1,\alpha}\Tr(W_{1,\alpha}\rho_{\alpha}) +  2\epsilon_{2,\alpha}\Tr(W_{2,\alpha}\rho_{\alpha}) + 4\epsilon_{1,\alpha}\epsilon_{2,\alpha}\Tr(W_{1,\alpha}W_{2,\alpha}) \right| \\
  \leqslant& 2\epsilon_{1,\alpha} \lVert W_{1,\alpha} \rVert_{\infty} \lVert \rho_{\alpha} \rVert_{1} + 2\epsilon_{2,\alpha} \lVert W_{2,\alpha} \rVert_{\infty} \lVert \rho_{\alpha} \rVert_{1} + 4\epsilon_{1,\alpha}\epsilon_{2,\alpha} \lVert W_{1,\alpha} \rVert_{1}\lVert W_{2,\alpha} \rVert_{\infty} \\
  \leqslant&  \epsilon_{1,\alpha} + \epsilon_{2,\alpha} + 2\epsilon_{1,\alpha}\epsilon_{2,\alpha} \leqslant \epsilon_{1} + \epsilon_{2} + 2\epsilon_{1}\epsilon_{2}, 
\end{split}
\end{equation}

and therefore:

\begin{equation}
\begin{split}
    \mathcal{E}_{K=2} =& \left| \mathcal{C}^{(K)}_{\ket{\psi'_1},\ket{\psi'_2}}(s)-\mathcal{C}^{(K)}_{\ket{\psi}}(s) \right| \leqslant \frac{2^{|s|}-1}{2^{|s|}} \left( \epsilon_{1} + \epsilon_{2} + 2\epsilon_{1}\epsilon_{2}  \right).
\end{split}
\end{equation}

Then, for the other odd prime $K$ in general, we have:

\begin{equation}
\begin{split}
    &\left | \Tr(\rho^K_{\alpha}) -  \frac{1}{K-1}\sum_{k=1}^{K-1}\Tr(\rho'_{1,\alpha}\rho'_{(1+k \Mod{K}), \alpha}\rho'_{(1+2k \Mod{K}),\alpha}\cdots\rho'_{(1+(K-1)k \Mod{K}), \alpha}) \right | \\
    =&\frac{1}{K-1} \left| \sum_{k=1}^{K-1}\left( \Tr(\rho^K_{\alpha})-\Tr(\rho'_{1,\alpha}\rho'_{(1+k \Mod{K}), \alpha}\rho'_{(1+2k \Mod{K}), \alpha}\cdots\rho'_{(1+(K-1)k \Mod{K}), \alpha}) \right)  \right| \\
    =&\frac{2}{K-1} \left| \sum_{k=1}^{\frac{K-1}{2}}\text{Re}\left(\Tr(\rho^K_{\alpha})-\Tr(\rho'_{1,\alpha}\rho'_{(1+k \Mod{K}), \alpha}\rho'_{(1+2k \Mod{K}), \alpha}\cdots\rho'_{(1+(K-1)k \Mod{K}), \alpha})\right) \right| \\
    \leqslant & \frac{2}{K-1}  \sum_{k=1}^{\frac{K-1}{2}}\left|\Tr(\rho^K_{\alpha})-\Tr(\rho'_{1,\alpha}\rho'_{(1+k \Mod{K}), \alpha}\rho'_{(1+2k \Mod{K}), \alpha}\cdots\rho'_{(1+(K-1)k \Mod{K}), \alpha}) \right|. \\
\end{split}
\end{equation}

Once we implement $\rho'_{k,\alpha}=\rho_{\alpha}+2\epsilon_{k,\alpha}W_{k,\alpha}$, we have:

\begin{equation}
\begin{split}
    &\left|\Tr(\rho^K_{\alpha})-\Tr(\rho'_{1,\alpha}\rho'_{(1+k \Mod{K}), \alpha}\rho'_{(1+2k \Mod{K}), \alpha}\cdots\rho'_{(1+(K-1)k \Mod{K}), \alpha}) \right| \\
    \leqslant& 2\sum_{k=1}^{K}\epsilon_{k,\alpha}\lVert W_{k,\alpha} \rVert_{\infty} +  2^2 \sum_{k<k'}\epsilon_{k,\alpha}\epsilon_{k',\alpha}\lVert W_{k,\alpha} \rVert_{\infty} \lVert W_{k',\alpha} \rVert_{\infty}+ \cdots + 2^K \left(\prod_{k=1}^{K}\epsilon_{k,\alpha}\right) \lVert W_{1,\alpha} \rVert_{1} \lVert W_{2,\alpha} \rVert_{\infty}  \cdots \lVert W_{K,\alpha} \rVert_{\infty} \\
    \leqslant& \sum_{k=1}^{K}\epsilon_k+\sum_{k<k'}\epsilon_k\epsilon_{k'}+\sum_{k<k'<k''}\epsilon_k\epsilon_{k'}\epsilon_{k''} +\cdots+2\prod_{k=1}^K\epsilon_k.
\end{split}
\end{equation}

Therefore:

\begin{equation}
\begin{split}
    &\left | \Tr(\rho^K_{\alpha}) -  \frac{1}{K-1}\sum_{k=1}^{K-1}\Tr(\rho'_{1,\alpha}\rho'_{(1+k \Mod{K}), \alpha}\rho'_{(1+2k \Mod{K}),\alpha}\cdots\rho'_{(1+(K-1)k \Mod{K}), \alpha}) \right | \\
    \leqslant & \sum_{k=1}^{K}\epsilon_k+\sum_{k<k'}\epsilon_k\epsilon_{k'}+\sum_{k<k'<k''}\epsilon_k\epsilon_{k'}\epsilon_{k''} +\cdots+2\prod_{k=1}^K\epsilon_k,
\end{split}
\end{equation}

and:

\begin{equation}
\begin{split}
    \mathcal{E} =& \left| \mathcal{C}^{(K)}_{\ket{\psi'_1},\ket{\psi'_2},\cdots,\ket{\psi'_K}}(s)-\mathcal{C}^{(K)}_{\ket{\psi}}(s) \right| \\
                \leqslant & \frac{2^{|s|}-1}{(K-1)2^{|s|}} \left( \sum_{k=1}^{K}\epsilon_k+\sum_{k<k'}\epsilon_k\epsilon_{k'}+\sum_{k<k'<k''}\epsilon_k\epsilon_{k'}\epsilon_{k''} +\cdots+2\prod_{k=1}^K\epsilon_k \right).
,\end{split}
\end{equation}

which also fits for $\mathcal{E}_{K=2}$. %(\textbf{Proposition \ref{prop3}} $\blacksquare$).

\end{proof}

Note that this error-bound, derived mostly from Schatten norm inequalities, is not sharp enough as it is very likely that all the equality cases in the inequalities cannot achieve simultaneously. 
From the numerical plots in Fig.~\ref{errorfig}, one can notice that the error scaling behaviour seems to be the case, but should be with smaller pre-factors. 

\subsection{GCE Circuit encoded in qubits} 
%\textcolor{blue}{Jerry's part}
In this section, we introduce a method for encoding the GCE circuit, which requires $K$-level qudits, using only qubits. All circuit-level numerical experiments presented in this work are conducted using this encoding method.

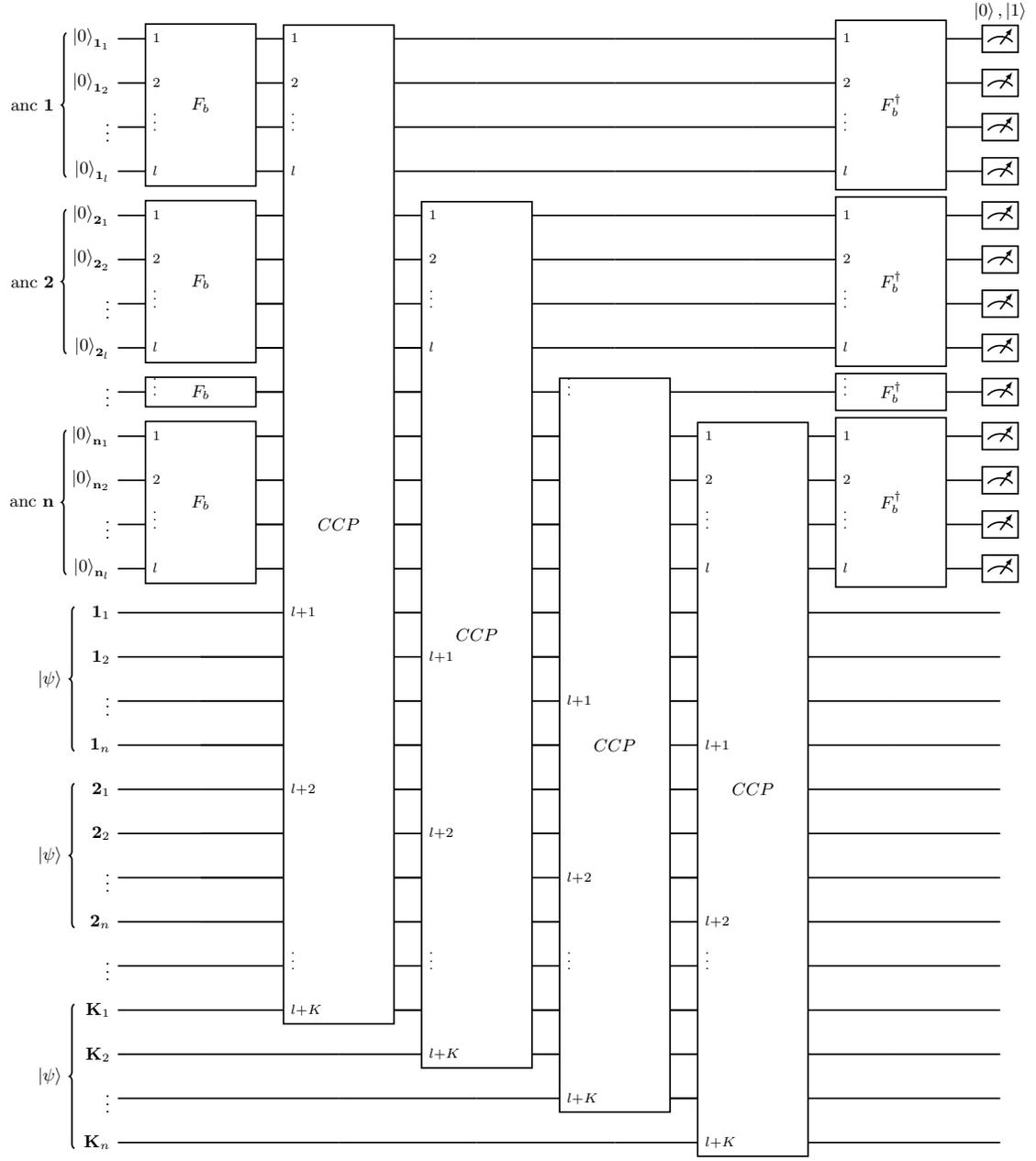
\begin{figure}  % 'h' places the figure approximately here, within the text
    \centering  % This centers the figure
    \begin{tikzpicture}
\node[scale=0.8]{
\begin{quantikz}[row sep = {0.8cm, between origins}]
% ancillas------------------------
% anc_1
\lstick[4]{$\text{anc} \ \mathbf{1}$} \quad
& \lstick{$\ket{0}_{\mathbf{1}_1}$} & \gate[4][2cm]{F_b}\gateinput{1} & \gate[23][2cm]{CCP}\gateinput{1} 
&\qw&\qw&\qw & \gate[4][2cm]{F^{\dag}_b}\gateinput{1} &\meter{$\ket{0},\ket{1}$} \\ 
& \lstick{$\ket{0}_{\mathbf{1}_2}$} & \gateinput{2} & \gateinput{2} &\qw&\qw&\qw & \gateinput{2} &\meter{} \\ 
& \lstick{$\mathbf{\vdots}$} & \gateinput{$\mathbf{\vdots}$} & \gateinput{$\mathbf{\vdots}$} &\qw&\qw&\qw & \gateinput{$\mathbf{\vdots}$} &\meter{} \\ 
& \lstick{$\ket{0}_{\mathbf{1}_l}$} & \gateinput{$l$} & \gateinput{$l$} &\qw&\qw&\qw & \gateinput{$l$} &\meter{} \\ 
% anc_2
\lstick[4]{$\text{anc} \ \mathbf{2}$} \quad
& \lstick{$\ket{0}_{\mathbf{2}_1}$} & \gate[4][2cm]{F_b}\gateinput{1} & \qw 
& \gate[20][2cm]{CCP}\gateinput{1} &\qw&\qw & 
\gate[4][2cm]{F^{\dag}_b}\gateinput{1} &\meter{} \\ 
& \lstick{$\ket{0}_{\mathbf{2}_2}$} & \gateinput{2} & \qw 
& \gateinput{2} &\qw&\qw & \gateinput{2} &\meter{} \\ 
& \lstick{$\mathbf{\vdots}$} & \gateinput{$\mathbf{\vdots}$} & \qw & \gateinput{$\mathbf{\vdots}$} &\qw&\qw & \gateinput{$\mathbf{\vdots}$} &\meter{} \\ 
& \lstick{$\ket{0}_{\mathbf{2}_l}$} & \gateinput{$l$} & \qw & \gateinput{$l$} \qw& \qw &\qw & \gateinput{$l$} &\meter{} \\ 
% anc_...
& \lstick{$\mathbf{\vdots}$} & \gate[1][2cm]{F_b}\gateinput{$\mathbf{\vdots}$} & \qw &\qw & 
\gate[17][2cm]{CCP}\gateinput{$
\mathbf{\vdots}$}& \qw & \gate[1][2cm]{F^{\dag}_b}\gateinput{$\mathbf{\vdots}$} &\meter{} \\ 
% anc_n
\lstick[4]{$\text{anc} \ \mathbf{n}$} \quad
& \lstick{$\ket{0}_{\mathbf{n}_1}$} & \gate[4][2cm]{F_b}\gateinput{1} & \qw & \qw & \qw & 
\gate[17][2cm]{CCP}\gateinput{1} & \gate[4][2cm]{F^{\dag}_b}\gateinput{1} &\meter{} \\ 
& \lstick{$\ket{0}_{\mathbf{n}_2}$} & \gateinput{2} & \qw & \qw & \qw & \gateinput{2} & \gateinput{2} &\meter{} \\ 
& \lstick{$\mathbf{\vdots}$} & \gateinput{$\mathbf{\vdots}$} & \qw & \qw & \qw & \gateinput{$\mathbf{\vdots}$} & \gateinput{$\mathbf{\vdots}$} &\meter{} \\ 
& \lstick{$\ket{0}_{\mathbf{n}_l}$} & \gateinput{$l$} & \qw & \qw & \qw & \gateinput{$l$} & \gateinput{$l$} &\meter{} \\ 
% state copies------------------------
% copy_1
\lstick[4]{$\ket{\psi}$} & \lstick{$\mathbf{1}_1$}          &\qw       &\gateinput{$l$+1} &\qw&\qw&\qw&\qw&\qw  \\
& \lstick{$\mathbf{1}_2$}          &\qw       &\qw                                      &\gateinput{$l$+1} &\qw &\qw &\qw&\qw \\
& \lstick{$\mathbf{\vdots}$}       &\qw       &\qw                                      &\qw     &\gateinput{$l$+1} &\qw&\qw&\qw\\
& \lstick{$\mathbf{1}_n$}          &\qw       &\qw                                      &\qw &\qw &\gateinput{$l$+1}&\qw&\qw \\
% copy_2
\lstick[4]{$\ket{\psi}$} & \lstick{$\mathbf{2}_1$}          &\qw       &\gateinput{$l$+2}                   &\qw &\qw &\qw &\qw&\qw \\
& \lstick{$\mathbf{2}_2$}          &\qw       &\qw                                      &\gateinput{$l$+2}  &\qw &\qw&\qw&\qw \\
& \lstick{$\mathbf{\vdots}$}       &\qw       &\qw                                      &\qw &\gateinput{$l$+2} &\qw&\qw&\qw \\
& \lstick{$\mathbf{2}_n$}          &\qw       &\qw                                      &\qw &\qw &\gateinput{$l$+2}&\qw&\qw \\
% copy_...
&\lstick{\vdots}          &\qw       &\gateinput{$\vdots$}            &\gateinput{$\vdots$} &\gateinput{$\vdots$} &\gateinput{$\vdots$}&\qw&\qw \\
% copy_K
\lstick[4]{$\ket{\psi}$} & \lstick{$\mathbf{K}_1$}          &\qw       &\gateinput{$l$+$K$}                 &\qw &\qw &\qw&\qw&\qw \\
& \lstick{$\mathbf{K}_2$}          &\qw       &\qw                                      & \gateinput{$l$+$K$}  &\qw &\qw&\qw&\qw \\
& \lstick{$\mathbf{\vdots}$}       &\qw       &\qw                                      &\qw &\gateinput{$l$+$K$} &\qw&\qw&\qw \\
& \lstick{$\mathbf{K}_n$}          &\qw       &\qw                                      &\qw &\qw &\gateinput{$l$+$K$}&\qw&\qw \\
\end{quantikz}};
\end{tikzpicture}
    \caption{Qubit-only cicuit for computing GCE with prime $K$. 
    Multiple copies of state $\psi$ are prepared and $n$ premutation tests are performed on the state copies in parallel with the help of $n$ group of ancilla qubits. Within each group, there are $l=\left\lceil \log_2 K \right\rceil$ qubits. 
    }  % Add your caption here
    \label{encoding_fig}  % This label is used to reference the figure
\end{figure}

The qubit-based circuit for efficiently computing GCE with prime $K$ is illustrated in Fig. \ref{encoding_fig}. 
In this encoding, each $K$-level ancilla qudit is represented by $l=\left\lceil \log_2 K \right\rceil$ qubits. 
The operator $F_b$ acts on the Hilbert space of these $l$ ancilla qubits. 
It performs Fourier transform only on the subspace spanned by the first $K$ basis states $\{ \ket{0\cdots00}, \ket{0\cdots01}, \cdots \ket{c_1c_2\cdots c_l} \}$, where $K = \sum_{i=1}^l 2^{l-i}c_i$ and $c_i \in \{0,1\}$,
but keep the rest of the space unchanged.
The matrix representation of $F_b$ in the computational basis is as follows:
\begin{equation} \label{F_b}
F_b = \begin{pmatrix}
  \frac{1}{\sqrt{K}}
  \begin{pmatrix}
  1 & 1 & 1 & \cdots & 1\\
  1 & \omega & \omega^2 & \cdots & \omega^{K-1}\\
  1 & \omega^2 & \omega^4 & \cdots & \omega^{2(K-1)}\\
  \vdots & \vdots & \vdots & \ddots & \vdots\\
  1 & \omega^{K-1} & \omega^{2(K-1)} & \cdots & \omega^{(K-1)(K-1)}
  \end{pmatrix}_{K\times K}
  & \rvline & \mathbf{0} \\
\hline
  \mathbf{0} & \rvline &
  \large\mathbb{I}_{2^l-K}
\end{pmatrix}, \quad \omega=e^{2\pi i/K}.
\end{equation}
In Eq.~\eqref{F_b}, the upper-left block of the matrix represents the $K$-level Fourier transform, while the lower-right block is an identity matrix of dimension $2^l-K$.

The controlled cyclic permutation ($CCP$) operates on the target state in the same manner as the multi-level controlled $D$, as shown in Eq.~\eqref{ccp}, 
in which $z = \sum_{i=1}^l 2^{l-i}c_i$.
\begin{equation}
    \begin{quantikz}[row sep = {0.5cm, between origins}]
    \lstick{$\ket{z}$}& \ctrl{1} &\qw \\
    \lstick{$\ket{\phi}$}&\gate{D} &\qw \rstick{$D^z\ket{\phi}$}\\
    \end{quantikz} \quad
    \rightarrow \quad
    \begin{quantikz}[row sep = {0.5cm, between origins}]
    \lstick[4]{\text{anc}}\
    & \lstick{$\ket{c_1}$}  & \gate[5][1.5cm]{CCP}\gateinput{} &\qw\\ 
    & \lstick{$\ket{c_2}$}  & \gateinput{} &\qw  \\ 
    & \lstick{$\mathbf{\vdots}$} & \gateinput{} &\qw \\ 
    & \lstick{$\ket{c_l}$} & \gateinput{} &\qw  \\ 
    & \lstick{$\ket{\phi}$} & \gateinput{} &\qw \rstick{$D^z\ket{\phi} \ \left(z = \sum_{i=1}^l 2^{l-i}c_i\right)$} \\ 
    \end{quantikz}
    \label{ccp}
\end{equation}

The elements of the matrix representation of $CCP$ operator in the computational basis is then:
\begin{equation}
\begin{split}
    CCP_{c'_1c'_2\cdots c'_lt'_1t'_2\cdots t'_K, c_1c_2\cdots c_lt_1t_2\cdots t_K}=&
    \bra{c'_1c'_2\cdots c'_l}\otimes \bra{t'_1t'_2\cdots t'_K} CCP \ket{c_1c_2\cdots c_l}\otimes \ket{t_1t_2\cdots t_K} \\
    =&\bra{c'_1c'_2\cdots c'_l} \ket{c_1c_2\cdots c_l}
     \bra{t'_1t'_2\cdots t'_K} \otimes D^z\ket{t_1t_2\cdots t_K}\\
    =& \prod_{i=1}^{l}\delta_{c'_ic_i} \prod_{j=1}^{K}\delta_{t'_jt_{j+z}},
\end{split}
\end{equation}
where $c_i$ denotes the $i$-th ancilla qubit and $t_j$ denotes the $j$-th qubit of the target state.

% \begin{equation}
% CCP = \begin{pmatrix}
%   A_1 & \rvline & \mathbf{0} & \rvline & \mathbf{0} & \rvline & \mathbf{0} & \rvline & \mathbf{0} &\\
% \hline
%   A_1 & \rvline & \mathbf{0} & \rvline & \mathbf{0} & \rvline & \mathbf{0} & \rvline & \mathbf{0} & \\
% \hline
%   A_1 & \rvline & \mathbf{0} & \rvline & \mathbf{0} & \rvline & \mathbf{0} & \rvline & \mathbf{0} &\\
% \hline
%   A_1 & \rvline & \mathbf{0} & \rvline & \mathbf{0} & \rvline & \mathbf{0} & \rvline & \mathbf{0} &\\
% \hline
%   A_1 & \rvline & \mathbf{0} & \rvline & \mathbf{0} & \rvline & \mathbf{0} & \rvline & \mathbf{0}
% \end{pmatrix}
% \end{equation}

\section{Mathematical Properties of GCE}

\subsection{Proof of Theorem \ref{theorem} \label{theoremproof}}

\subsubsection{Proof of Theorem \ref{theorem}.1}

\begin{proof}

Here we only give brief information about the proof as it is almost the same as the one in Ref.~\cite{beckey2021computable} (Supplementary Information I) for $K=2$. 
The only change is that we need to make sure that $\Tr(\rho^K)$ is a convex function in general for $\forall K>1$, i.e. $\Tr\left((p\rho_1+(1-p)\rho_2)^K\right) \leqslant p\Tr(\rho^K_1) + (1-p)\Tr(\rho^K_2)$ for $p\in[0,1]$. 
This statement is true because of Ref.~\cite{hu2006generalized} (Lemma 1), where the authors found that $\left(\Tr(\rho^r)\right)^s$ is a convex function of the density operator $\rho$ for $r\geqslant 1$ and $rs \geqslant 1$, which perfectly includes the case for $\Tr(\rho^K)$ when $K\geqslant1$. %(\textbf{Theorem \ref{theorem} 1)} $\blacksquare$).

\end{proof}

\subsubsection{Proof of Theorem \ref{theorem}.2}

\begin{proof}

This proof is also similar in Ref.~\cite{beckey2021computable}, where we can easily know that if $\rho$ is completely separable, then $\rho_{\alpha}$ is pure always. 
Therefore, $\Tr(\rho_{\alpha}^K)=1$ for $\forall K>1$, which outputs 0 for GCE in the end. %(\textbf{Theorem \ref{theorem} 2)} $\blacksquare$).

\end{proof}

\subsubsection{Proof of Theorem \ref{theorem}.3}

\begin{proof}

To have the continuity, as:

\begin{equation}
    \left|\mathcal{C}^{(K)}_{\ket{\psi}}(s)-\mathcal{C}^{(K)}_{\ket{\psi'}}(s)\right| = \frac{1}{(K-1)2^{|s|}}\left| \sum_{\alpha\in\mathcal{P}(s)} \left(\Tr(\rho_{\alpha}^K) - \Tr({\rho'_{\alpha}}^K) \right) \right| \leqslant \frac{1}{2^{|s|}} \sum_{\alpha\in\mathcal{P}(s)} \frac{1}{K-1}\left| \Tr(\rho_{\alpha}^K) - \Tr({\rho'_{\alpha}}^K)  \right|.
\end{equation}

Now, to find the upper-bound of $\left| \Tr(\rho_{\alpha}^K) - \Tr({\rho'_{\alpha}}^K)  \right|$, we remind the continuity of Tsallis entropy in Ref.~\cite{raggio1995properties} (Lemma 2):

\begin{equation}
    \left|T_K(\rho) - T_K(\rho') \right| = \frac{1}{K-1}\left| \Tr(\rho^K)-\Tr({\rho'}^K) \right| \leqslant \frac{K}{K-1} \lVert \rho - \rho' \rVert_{1} \leqslant \frac{2K}{K-1}\epsilon.
\end{equation}

Therefore, we have:

\begin{equation}
    \left|\mathcal{C}^{(K)}_{\ket{\psi}}(s)-\mathcal{C}^{(K)}_{\psi'}(s)\right| \leqslant \frac{2K}{K-1}\epsilon.
\end{equation}

%(\textbf{Theorem \ref{theorem} 3)} $\blacksquare$).

\end{proof}

Note that there may exist a sharper continuity bound from the continuity of Tsallis entropy derived in Ref.~\cite{hanson2017tight} (Theorem 3.1), which illustrates that:

\begin{equation}
    \left|T_K(\rho) - T_K(\rho') \right| = \frac{1}{K-1}\left| \Tr(\rho^K)-\Tr({\rho'}^K) \right| \leqslant \left\{
    \begin{aligned}
    &\frac{1}{K-1}\left( 1-(1-\epsilon)^K-(d-1)^{1-K}\epsilon^K \right), \ \ &(\epsilon < 1-\frac{1}{d}) \\
    &\frac{1}{K-1}\left( 1-d^{1-K} \right), &(\epsilon \geqslant 1-\frac{1}{d})
    \end{aligned}
    \right. .
\end{equation}

where $d$ is the Hilbert space dimensions of $\rho$ (or $\rho'$) and it will vary for different subsystem $\alpha$.

\subsubsection{Proof of Theorem \ref{theorem}.4}

\begin{proof}

We denote $\alpha\in\mathcal{P}(S)$ and $\alpha'\in\mathcal{P}(S\backslash\{n_0\})$. This theorem can be simplified to prove $\sum_{\alpha}\Tr(\rho_{\alpha}^K)=2\sum_{\alpha'}\Tr(\rho_{\alpha'}^K)$. 
Note that $\Tr(\rho^K_{\alpha}) = \Tr(\rho^K_{\bar{\alpha}})$ for any pure density matrix $\rho$ and $\bar{\alpha}\in S\backslash \alpha$, which can be easily seen from Schmidt decomposition. 
Therefore:

\begin{equation}
    \sum_{\alpha\in\mathcal{P}(s)}\Tr(\rho^K_{\alpha}) = \sum_{\alpha'\in\mathcal{P}(s)\backslash\{n_0\}}\left(\Tr(\rho^K_{\alpha'}) + \Tr(\rho^K_{\alpha'\cup\{n_0\}})\right)=2\sum_{\alpha'\in\mathcal{P}(s)\backslash\{n_0\}}\Tr(\rho^K_{\alpha'}).
\end{equation}

%(\textbf{Theorem \ref{theorem} 4)} $\blacksquare$).

\end{proof}

\subsection{Discussions on Conjectures \ref{conjecture} \label{discussconjecture}}

In this section, we would like to provide more comments and discussions on Conjectures \ref{conjecture}, which may be helpful for the further analytical proof, or find a counter-example for it.

\subsubsection{Similar methods in Ref.~\cite{beckey2021computable}}

The difficulty of the proof is from the pre-factor $1/2^{|s|}$, which can vary for different $s$. 
In Ref.~\cite{beckey2021computable} (Supplementary Information I), the authors provide a very concise proof for Conjectures \ref{conjecture} when $K=2$. 
They estimate their GCE in an elegant way, where they found:

\begin{equation}
    \mathcal{C}^{(2)}_{\ket{\psi}}(s)= 1 - \sum_{\mathbf{z} \in \mathcal{Z}_{0}(s)}\mathfrak{p}(\mathbf{z}),
\end{equation}

where:

\begin{equation}
    \mathfrak{p}(\mathbf{z}) = \frac{1}{2^n}\bra{\Psi}\prod_{j}(\mathbb{I}_j+(-1)^{z_j}S_j)\ket{\Psi},
\label{CEpz}
\end{equation}

and $\mathbf{z} \in\mathcal{Z}_{0}(s)$ denotes that $\mathbf{z}$ is a bitstring s.t. all bits in the label set $s$ should be 0. 
They use the parallelized SWAP test to directly estimate the $\mathfrak{p}(\mathbf{z})$ from the probability distribution of the ancillas, which is always equal to or larger than 0. 
Then, the proof is simple and straightforward as the pre-factor is dropped. 
However, finding a circuit that directly replaces $S$ into $D$ in Eq.~\eqref{CEpz} is hard as $D$ is not Hermitian for $K\geqslant 3$. 
That is why we use another way to estimate GCE but limited for prime $K$ as shown in Proposition~\ref{prop1}. 

However, for mathematical perspective, we can still define a quantity:

\begin{equation}
    \mathfrak{q}(\mathbf{z}) = \frac{1}{2^n}\bra{\Psi}\prod_{j}(\mathbb{I}_j+(-1)^{z_j}D_j)\ket{\Psi} = \frac{1}{2^n}\sum_{\alpha\in\mathcal{P}(s)}(-1)^{\sum_{x\in\alpha}z_x}\Tr(\rho_{\alpha}^K),
\end{equation}

where $\ket{\Psi}=\ket{\psi}^{\otimes K}$. 
Then if $\mathfrak{q}(\mathbf{z})\geqslant0$, both conjectures for integer $K\geqslant2$ hold by following the similar proof in Ref.~\cite{beckey2021computable}. 
Importantly, for simplicity, here we suppose $D=D^{(1)}$ as in this way the SWAP trick holds for not only prime $K$ but also composite $K$. 
Also, instead of higher-level ancillas, here we consider $\mathbf{z}\in\{0,1\}^{n}$.

We start from proving a lemma for $\mathfrak{q}(\mathbf{z})$:

\textbf{Lemma 2:} \textit{If there are odd number of '1's in $\mathbf{z^o}=z^o_1z^o_2\cdots z^o_n$, then $\mathfrak{q}(\mathbf{z^o})=0$.}

\textit{\textbf{Proof of Lemma 2:}} 

\begin{proof}
Since:

\begin{equation}
    \begin{split}
        \mathfrak{q}(\mathbf{z^o}) &= \frac{1}{2^n}\bra{\Psi}\prod_{j}(\mathbb{I}_j+(-1)^{z^o_j}D_j)\ket{\Psi} \\
        &= \frac{1}{2^n} (1+(-1)^{z^o_1}\Tr(\rho^K_1)+(-1)^{z^o_2}\Tr(\rho^K_2)+\cdots+(-1)^{z^o_1+z^o_2+\cdots+z^o_n}\Tr(\rho^K)).
    \end{split}
\end{equation}

If there are odd number of '1's in $\mathbf{z^o}$, consider:

\begin{equation}
    (-1)^{\sum_{\alpha}z^o_{\alpha}}\Tr(\rho^K_{\alpha}) = (-1)^{\sum_{\alpha}z^o_{\alpha}}\Tr(\rho^K_{\bar{\alpha}}) = -(-1)^{\sum_{\bar{\alpha}}z^o_{\bar{\alpha}}}\Tr(\rho^K_{\bar{\alpha}}).
\end{equation}

Therefore, by paring the terms and their corresponding complimentary, we have $\mathfrak{q}(\mathbf{z^o})=0$.

\end{proof}

Also, since $\mathfrak{q}(\mathbf{0})\geqslant0$, we can only consider the case where $\mathbf{z}$ has non-zero even number of '1's. 
In the following part, we show several proofs of small $n$, and then illustrate why it is hard to generalize.

\begin{itemize}
    \item When $n=1$, the proof is trivial.
    \item When $n=2$, the only non-trivial case is:
    \begin{equation}
        \mathfrak{q}(11) = \frac{1}{4}(2-\Tr(\rho^{K}_1)-\Tr(\rho^{K}_2)),
    \end{equation}
    which is obviously $\geqslant0$.
    \item When $n=3$, w.l.o.g. we need to consider the case:
    \begin{equation}
    \begin{split}
        \mathfrak{q}(110) =& \frac{1}{8}(2-\Tr(\rho^{K}_1)-\Tr(\rho^{K}_2)+\Tr(\rho^{K}_3)+\Tr(\rho^{K}_{12})-\Tr(\rho^{K}_{13})-\Tr(\rho^{K}_{23}))  \\
        =& \frac{1}{4}(1-\Tr(\rho^K_1)-\Tr(\rho^K_2)+\Tr(\rho^K_{12})).
    \end{split}
    \end{equation}
    Because Tsallis entropy has the property of subadditivity for $K>1$ \cite{audenaert2007subadditivity}, i.e.,
    \begin{equation}
    \begin{split}
        &T_K(\rho_{AB}) \leqslant T_K(\rho_{A})+T_K(\rho_B) \\
    \rightarrow&\Tr(\rho^K_A)+\Tr(\rho^K_B)\leqslant1+\Tr(\rho^K_{AB}).
    \end{split}
    \end{equation}
    Then $\mathfrak{q}(110)\geqslant0$. 
    \item When $n=4$, w.l.o.g. we need to consider:
    \begin{equation}
        \mathfrak{q}(1100) = \frac{1}{8}\left( \left(1+\Tr(\rho^K_{12})-\Tr(\rho^K_1)-\Tr(\rho^K_2)\right) + \left(\Tr(\rho^K_{124})+\Tr(\rho^K_{4})-\Tr(\rho^K_{14})-\Tr(\rho^K_{24})\right) \right),
    \end{equation}
    and:
    \begin{equation}
        \mathfrak{q}(1111) = \frac{1}{8}\left( \left(1+\Tr(\rho^K_{12})-\Tr(\rho^K_1)-\Tr(\rho^K_2)\right) - \left(\Tr(\rho^K_{124})+\Tr(\rho^K_{4})-\Tr(\rho^K_{14})-\Tr(\rho^K_{24})\right) \right).
    \end{equation} 
    This is where the difficulty emerges as the first inner bracket denotes the subadditivity (SA) of Tsallis entropy and the second inner bracket:
    \begin{equation}
    \begin{split}
        &\Tr(\rho^K_{ABC})+\Tr(\rho^K_{C})-\Tr(\rho^K_{AC})-\Tr(\rho^K_{BC}) \\
        \rightarrow & T_K(\rho_{AC})+T_K(\rho_{BC}) - T_K(\rho_{ABC}) - T_K(\rho_{C})
    \end{split}
    \end{equation}
    has the form of strong subadditivity (SSA), which does not in general hold positively or negatively for Tsallis entropy~\cite{petz2015inequalities}. 
\end{itemize}

From the above example, we see that if $\mathfrak{q}(\mathbf{z})\geqslant0$ holds, there may exist a sharper inequality than the SA. In fact, if we pick:

\begin{equation}
    \mathfrak{q}(1_B1_C0_{A_1}0_{A_2}\cdots0_{A_{n-2}}) = \frac{1}{2^n}\sum_{\alpha_A\in \mathcal{P}(\{A_1,\cdots,A_{n-2}\})}\left(\Tr(\rho^K_{\alpha_A BC})+\Tr(\rho^K_{\alpha_A})-\Tr(\rho^K_{\alpha_A B}) - \Tr(\rho^K_{\alpha_A C})\right),
\end{equation}

we have the not-so-strong subadditivity (NSSSA) that may hold, which is the sum over all possible SSA w.r.t. the subsets of '0'-labels.

\begin{figure*}
    \centering
    \includegraphics[width=0.75\linewidth]{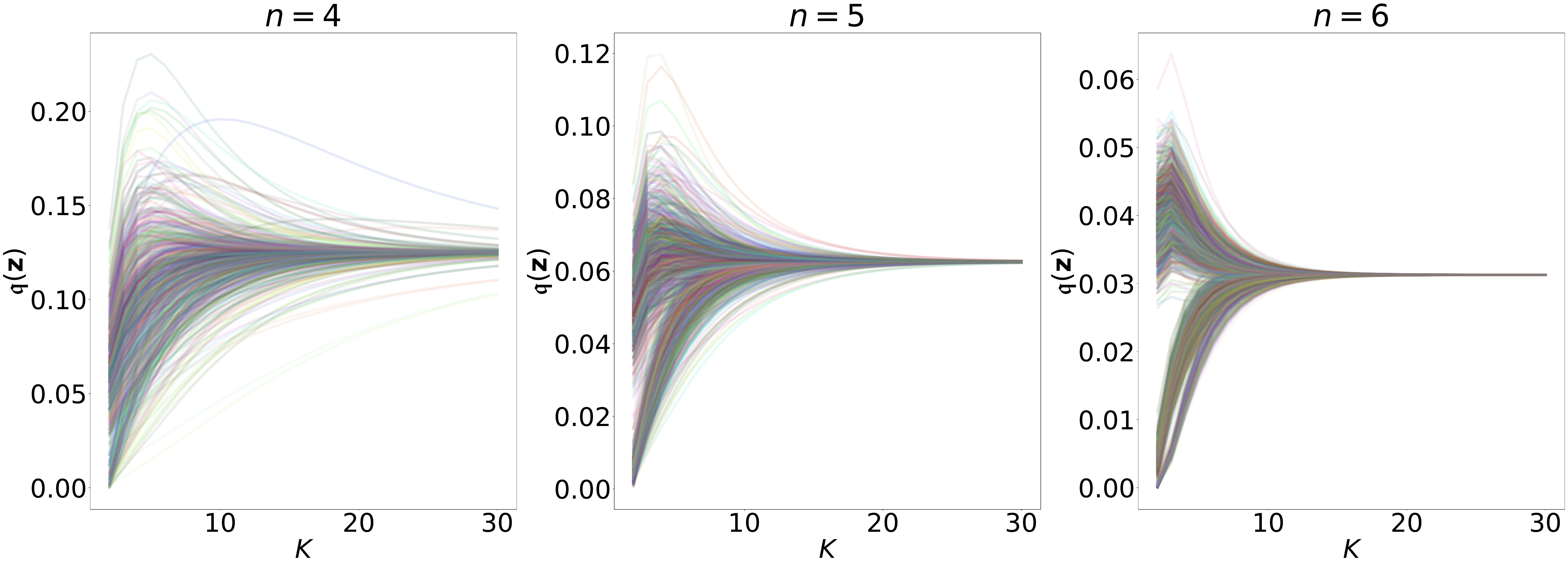}
    \caption{Numerical simulations for $\mathfrak{q}(\mathbf{z})$ under non-trivial $\mathbf{z}$ and $n=4,5,6$. 
    Here, each Haar random statevector is initialized at the beginning and then we plot its corresponding $\mathfrak{q}(\mathbf{z})$ under increasing $K$ and a certain non-trivial $\mathbf{z}$. 
    There is no counter-example yet s.t. $\mathfrak{q}(\mathbf{z})<0$.}
    \label{qzfig}
\end{figure*}

We also provide some numerics for $\mathfrak{q}(\mathbf{z})$, as shown in Fig.~\ref{qzfig}. 
For each plotted line, a Haar random statevector and a certain non-trivial $\mathbf{z}$ are picked at the beginning and then its $\mathfrak{q}(\mathbf{z})$ values with the evolution of $K$ are plotted. 
We found that there is no single counter-example that gives $\mathfrak{q}(\mathbf{z})<0$. 
Also, $\mathfrak{q}(\mathbf{z})$ will go asymptotically to $2^{1-n}$ and this can be directly seen from the mathematical form of $\mathfrak{q}(\mathbf{z})$.

Below we provide some other possible techniques for proving $\mathfrak{q}(\mathbf{z})\geqslant0$.

\begin{enumerate}
    \item \textit{Mathematical Induction.} As the case of $K=2$ holds, we can suppose the case of $K$ also hold and try to prove the case of $K+1$. However, there are no clear mathematical connections between $K$ and $K+1$.
    \item \textit{Eigenspace of $\prod_{j}(\mathbb{I}_j+(-1)^{z_j}D_j)$:} Suppose $\prod_{j}(\mathbb{I}_j+(-1)^{z_j}D_j)=M$, then $\mathfrak{q}(\mathbf{z})\sim\Tr(M\rho^{\otimes K})$. 
    Write $\rho^{\otimes K}=\sum_{i}\lambda_i\ket{\psi_i}\bra{\psi_i}$ and $M=\sum_{i}\sigma_i\ket{\phi_i}\bra{\phi_i}$, therefore $\mathfrak{q}(\mathbf{z})\sim \sum_{i,j}\sigma_i\lambda_j\left|\braket{\phi_i}{\psi_j}\right|^2$. 
    Since $\lambda\geqslant0$ and $\mathfrak{q}(\mathbf{z})\in\mathbb{R}$, an optimistic assumption to prove $\mathfrak{q}(\mathbf{z})\geqslant0$ is that $\text{Re}(\sigma)\geqslant0$. 
    This is true for $K=2$ but not for $K\geqslant3$. However, both $M$ and $\rho^{\otimes K}$ are not any general diagonalizable matrices and they have intrinsic symmetries. 
    Writing $\rho^{\otimes K}$ into Dicke basis might be more helpful to find the proof~\cite{dicke1954coherence}.
    \item \textit{Sum of squares (SOS):} We can symbolize the statevector $\ket{\psi}$:
    \begin{equation}
        [a_0+ib_0, a_1+ib_1, \cdots a_{2^n-1}+ib_{2^n-1}],
    \end{equation}
    where $a$ and $b$ denote the real and imaginary part of each statevector element, respectively. 
    After calculations, we should have $\mathfrak{q}(\mathbf{z})$ in a cumbersome polynomial. 
    We would like to see if this polynomial is an SOS. 
    However, as GCE is invariant up to local unitaries, we can change the basis of $\ket{\psi}$ via local unitaries without modifying the values of $\mathfrak{q}(\mathbf{z})$. 
    Therefore, with the help of the generalized Schmidt decomposition in Ref.~\cite{carteret2000multipartite} (Theorem 1), several elements in $\ket{\psi}$ can be simplified to 0. 
    Due to the exponentially large system with qubit number $n$, we tried the real statevector $\ket{\psi}$ for $\mathfrak{q}(0011)$ ($n=4$) and $K=3$, and we used Mosek SDP solver~\cite{aps2024mosek} and SumOfSquares~\cite{weisser2019polynomial,legat2017sumofsquares} package in Julia for the SOS polynomial optimizations. 
    The optimization gives that $\mathfrak{q}(0011)+\varepsilon=\text{SOS}$, where $\varepsilon$ is a very small positive number that varies device by device, depending on the device precision. 
    The smallest $\varepsilon$ we found is 0.009901732. 
    Hence, whether $\mathfrak{q}(\mathbf{z})$ is an SOS is still unknown, and we are inclined to believe that this polynomial is not an SOS.
\end{enumerate}

\subsubsection{More Comments on Conjecture \ref{conjecture}.1 and Proof of Proposition~\ref{prop5} \label{NSSSAprop}}

In this section we provide more discussions on Conjecture \ref{conjecture}.1. 
Conjecture \ref{conjecture}.1 can be transformed into a simpler formulation:

\begin{equation}
    \mathcal{C}^{(K)}_{\ket{\psi}}(s')\leqslant\mathcal{C}^{(K)}_{\ket{\psi}}(s), \ \ \textit{where $s=s'\cup\{n_0\}$ for single label $n_0\in S\backslash s'$.}
    \label{simp1.3}
\end{equation}

Starting from simple cases, we list the checklist in Table.~\ref{table:conj1}, where we present if Eq.~\eqref{simp1.3} has been proved or not for any $K>1$. 
Due to the transitivity of the inequality, we can let $s'=\{1,2,\cdots,|s'|\}$ and $s=\{1,2,\cdots,|s'|,|s|\}$. 

\begin{itemize}
    \item If $|s|=2$, then Eq.~\eqref{simp1.3} goes to SA of Tsallis entropy, which holds for any $K>1$.
    \item If $|s|=n$, i.e., $s=S$, the equality holds for Eq.~\eqref{simp1.3} due to Theorems \ref{theorem}.4. 
\end{itemize}

\begin{table}[h]
\centering
\caption{This checklist shows the information that if Eq.~\eqref{simp1.3} has been proved or not in this work, for certain $n$ and $|s|$ (and any real $K>1$).}
\renewcommand{\arraystretch}{2}
\setlength{\tabcolsep}{15pt}
\scalebox{1}{\begin{tabular}{|c|c|c|c|c|c|c|}
\hline
    & $n=1$   & $n=2$   & $n=3$   & $n=4$   & $n=5$   & $\cdots$ \\ \hline
$|s|=1$   & N/A  & N/A  & N/A  & N/A  & N/A  & N/A  \\ \hline
$|s|=2$   & N/A & \textcolor{magenta}{\checkmark}  & \textcolor{magenta}{\checkmark}  & \textcolor{magenta}{\checkmark}  & \textcolor{magenta}{\checkmark}  & \textcolor{magenta}{\checkmark}  \\ \hline
$|s|=3$   & N/A & N/A & \textcolor{teal}{\checkmark}  & \textcolor{orange}{\textbf{?}}  & \textcolor{orange}{\textbf{?}}  & \textcolor{orange}{\textbf{?}}  \\ \hline
$|s|=4$   & N/A & N/A & N/A & \textcolor{teal}{\checkmark}  & \textcolor{orange}{\textbf{?}}  & \textcolor{orange}{\textbf{?}}  \\ \hline
$|s|=5$   & N/A & N/A & N/A & N/A & \textcolor{teal}{\checkmark}  & \textcolor{orange}{\textbf{?}}  \\ \hline
$\vdots$ & N/A & N/A & N/A & N/A & N/A & \textcolor{teal}{\checkmark}  \\ \hline
\end{tabular}}
\label{table:conj1}
\end{table}

The other cases are hard to prove and it also relates to the NSSSA form of Tsallis entropy, which goes to the proof of Proposition~\ref{prop5}:

\begin{proof}

Suppose $\bar{s'}=S\backslash s'=\{|s|, |s|+1,\cdots,n\}$ and $\bar{s}=S\backslash s = \{|s|+1, |s|+2, \cdots, n\}$, we have the sum over SSA:

\begin{equation}
\begin{split}
    &\sum_{\alpha'\in \mathcal{P}(s')}T_K(\rho_{\alpha'\cup\{|s|\}\cup\bar{s}})+T_K(\rho_{\alpha'})-T_K(\rho_{\alpha'\cup\{|s|\}})-T_K(\rho_{\alpha'\cup\bar{s}}) \\
    =&\sum_{\alpha'\in \mathcal{P}(s')}T_K(\rho_{\alpha'\cup\bar{s'}})+T_K(\rho_{\alpha'})-T_K(\rho_{\alpha'\cup\{|s|\}})-T_K(\rho_{\alpha'\cup\bar{s}}) \\
    =&\sum_{\alpha'\in \mathcal{P}(s')} T_K(\rho_{\alpha'})+T_K(\rho_{\alpha'})-T_K(\rho_{\alpha'\cup\{|s|\}})-T_K(\rho_{\alpha'\cup\{|s|\}}) \\
    =&2\sum_{\alpha'\in\mathcal{P}(s')}\left(\Tr(\rho^K_{\alpha'\cup\{|s|\}})-\Tr(\rho^K_{\alpha'})\right) \\
    \propto& \mathcal{C}^{(K)}_{\ket{\psi}}(s') - \mathcal{C}^{(K)}_{\ket{\psi}}(s).
\end{split}
\end{equation}

\end{proof}

Moreover, majorization may be another way to prove this. 
We can place all the eigenvalues of $\rho_{\alpha'\cup\{|s|\}}$ and $\rho_{\alpha'}$ non-increasingly into two lists $\boldsymbol{\lambda}_{\alpha'\cup\{|s|\}}$ and $\boldsymbol{\lambda}_{\alpha'}$, respectively. 
Note that there are of course fewer eigenvalues in $\boldsymbol{\lambda}_{\alpha'}$ and we fill in zeros to align the lengths of $\boldsymbol{\lambda}_{\alpha'\cup\{|s|\}}$ and $\boldsymbol{\lambda}_{\alpha'}$. 
As half of the elements in $\boldsymbol{\lambda}_{\alpha'}$ are 0 and its maximum element is 1. 
It is very likely that $\boldsymbol{\lambda}_{\alpha'}\succ\boldsymbol{\lambda}_{\alpha'\cup\{|s|\}}$. 
If this holds, then a Schur convex function $f:\mathbb{R}^n\rightarrow\mathbb{R}$ will give $f(\boldsymbol{\lambda}_{\alpha'\cup\{|s|\}})\leqslant f(\boldsymbol{\lambda}_{\alpha'})$~\cite{nielsen1999conditions}. 
As $\sum\lambda^K$ is convex for $K>1$, then the conjecture becomes true. 
However, the rigorous proof for the majorization relation between $\boldsymbol{\lambda}_{\alpha'}$ and $\boldsymbol{\lambda}_{\alpha'\cup\{|s|\}}$ is still unknown.

\subsection{Analytical Details for GCE Examples \label{examples}}

In the main text, we illustrate two examples of GCE.
Here we provide more analytical details.

Firstly, we consider the spin-squeezed states. 
The spin-squeezed states we considered in this work are the states that are prepared by evolving a coherent spin state under the one-axis twisting Hamiltonian operator ($H_{OAT}=\chi \hat{S}^2_z$)~\cite{kitagawa1993squeezed} with time-evolution $t$. 
Then the final state becomes~\cite{guo2023detecting}:

\begin{equation}
    \ket{\Phi(\mu)}=\frac{1}{2^{\frac{n}{2}}}\sum_{k=0}^n \sqrt{\binom{n}{k}}e^{-i\frac{(n/2-k)^2\mu}{2}}\ket{D(n,k)},
\end{equation}

where $\ket{D(n,k)}$ is the $n$-qubit Dicke state with Hamming weight $w$ is $k$~\cite{dicke1954coherence}:

\begin{equation}
    \ket{D(n,k)} = \frac{1}{\sqrt{\binom{n}{k}}}\sum_{\substack{x\in\{0,1\}^{\otimes n}\\  w(x)=k}}\ket{x},
\end{equation}

and $\mu=2\chi t$. 
Then we have the density matrix of the pure state $\ket{\Phi(\mu)}$ as:

\begin{equation}
\begin{split}
    \rho^{\mu} = \ket{\Phi(\mu)}\bra{\Phi(\mu)}=&\frac{1}{2^n}\sum^n_{k,l=0}\sqrt{\binom{n}{k}\binom{n}{l}}\exp{i\frac{\mu}{2}(k-l)(n-l-k)}\ket{D(n,k)}\bra{D(n,l)} \\
    =& \sum_{k,l=0}^{n}c_{k,l}\ket{D(n,k)}\bra{D(n,l)},
\end{split}
\end{equation}

where:

\begin{equation}
    c_{k,l} = \frac{1}{2^n}\sqrt{\binom{n}{k}\binom{n}{l}}\exp{i\frac{\mu}{2}(k-l)(n-l-k)}.
\end{equation}

To get the reduced density matrix of this symmetric state on the subsystem $\alpha$, in~\cite{aloy2021quantum} (Theorem 1), we have:

\begin{equation}
    \rho^{\mu}_{\alpha}=\sum_{k_{\alpha},l_{\alpha}=0}^{|\alpha|}c_{k_{\alpha},l_{\alpha}}\ket{D(|\alpha|,k_{\alpha})}\bra{D(|\alpha|,l_{\alpha})},
\end{equation}

and:

\begin{equation}
\begin{split}
c_{k_{\alpha},l_{\alpha}}=&\sum_{k,l=0}^{n}c_{k,l}\sum_{\kappa=0}^{n-|\alpha|}\frac{(n-|\alpha|)!}{\kappa!(n-|\alpha|-\kappa)!}\sqrt{\frac{ \frac{|\alpha|!}{k_{\alpha}!(|\alpha|-k_{\alpha})!} \frac{|\alpha|!}{l_{\alpha}!(|\alpha|-l_{\alpha})!}  }{ \frac{n!}{k!(n-k)!} \frac{n!}{l!(n-l)!}  }}\delta(\kappa+k_\alpha-k)\delta(\kappa+l_\alpha-l) \\
=&\sum_{k,l=0}^{n}c_{k,l}\sum_{\kappa=0}^{n-|\alpha|}\frac{(n-|\alpha|)!}{\kappa!(n-|\alpha|-\kappa)!}\frac{|\alpha|!}{n!}\sqrt{\frac{k!l!(n-k)!(n-l)!}{k_{\alpha}!l_{\alpha}!(|\alpha|-k_{\alpha})!(|\alpha|-l_{\alpha})!}}\delta(\kappa+k_\alpha-k)\delta(\kappa+l_\alpha-l).
\end{split}
\end{equation}

In this way, we are able to represent $\rho^{\mu}_{\alpha}$ in a $(|\alpha|+1)\times(|\alpha|+1)$ matrix (instead of $2^{|\alpha|}\times 2^{|\alpha|}$ in the computational basis). 
By diagonalizing each $\rho^{\mu}_{\alpha}$, one can acquire their eigenvalues to estimate the corresponding GCE finally.

Secondly, we consider the GCE comparison between $\ket{GHZ}$ and $\ket{W}$ states.
Starting from $n$-qubit $\ket{GHZ}$ state:

\begin{equation}
    \ket{GHZ} = \frac{1}{\sqrt{2}}(\ket{00\cdots0}+\ket{11\cdots1}),
\end{equation}

then the pure $\ket{GHZ}$ density matrix is:

\begin{equation}
    \rho_{\ket{GHZ}} = \frac{1}{2}(\ket{00\cdots0}\bra{00\cdots0}+\ket{00\cdots0}\bra{11\cdots1}+\ket{11\cdots1}\bra{00\cdots0}+\ket{11\cdots1}\bra{11\cdots1}).
\end{equation}

Then, by tracing out the system $\bar{\alpha}$, we have:

\begin{equation}
    \rho_{\ket{GHZ},\alpha} = \Tr_{\bar{\alpha}}(\rho_{GHZ}) = \frac{1}{2} (\ket{00\cdots0}\bra{00\cdots0}  + \ket{11\cdots1}\bra{11\cdots1} ) \ \ \ \ \ \text{($|\alpha|$ qubits)}.
\end{equation}

Then, except for $\alpha=\varnothing$ or $\alpha=S$, any other reduced system will have two non-trivial eigenvalues of $\frac{1}{2}$ and in this case $\Tr(\rho_{\alpha}^K)=\frac{1}{2^{K-1}}$. Then:

\begin{equation}
    \mathcal{C}^{(K)}_{\ket{GHZ}}(s) = 1 - \frac{1}{2^{|s|-\delta(|s|,n)}}\left(1+\frac{2^{|s|-\delta(|s|,n)}-1}{2^{K-1}}\right).
\end{equation}

Next, consider the $n$-qubit $\ket{W}$ state:

\begin{equation}
    \ket{W} = \frac{1}{\sqrt{n}}\left( \ket{100\cdots0}+\ket{010\cdots0}+\ket{001\cdots0}+\cdots+\ket{000\cdots1} \right),
\end{equation}

and its pure state density matrix:

\begin{equation}
    \rho_{\ket{W}} = \frac{1}{n}(\ket{100\cdots0}\bra{100\cdots0}+\ket{100\cdots0}\bra{010\cdots0}+\cdots+\ket{000\cdots1}\bra{000\cdots1}).
\end{equation}

Then, by tracing out the system $\bar{\alpha}$, we have:

\begin{equation}
    \rho_{\ket{W},\alpha}=\Tr_{\bar{\alpha}}(\rho_{\ket{W}}) = \frac{n-|\alpha|}{n}\ket{000\cdots0}\bra{000\cdots0}+\frac{|\alpha|}{n}\ket{W_{|\alpha|}}\bra{W_{|\alpha|}}.
\end{equation}

Here, $\ket{W_{|\alpha|}}$ denotes the $|\alpha|$-qubit $\ket{W}$ state. 
Therefore, the two non-trivial eigenstates for $\rho_{\ket{W},\alpha}$ are $\frac{n-|\alpha|}{n}$ and $\frac{|\alpha|}{n}$, which gives $\Tr(\rho^K_{\ket{W},\alpha}) = \left(\frac{n-|\alpha|}{n}\right)^K+\left(\frac{|\alpha|}{n}\right)^K$ and finally:

\begin{equation}
    \mathcal{C}^{(K)}_{\ket{W}}(s) = 1 - \frac{1}{2^{|s|}n^K} \sum_{j=0}^{|s|}\left(\binom{|s|}{j}(n-j)^K+\binom{|s|}{j}j^K \right).
\end{equation}

%(\textbf{Proposition \ref{prop4}} $\blacksquare$).

\end{document}